\documentclass[12pt]{article}
\usepackage[utf8x]{inputenc}
\usepackage{arydshln}
\usepackage{setspace}

\usepackage{amssymb}
\usepackage{geometry}
\usepackage{amsthm}
\usepackage{envmath}
\usepackage{subfigure}
\usepackage{amssymb}
\usepackage{graphicx}

\usepackage{amsmath}
\usepackage{lscape}

\usepackage[dvipsnames]{xcolor}

\usepackage{bm}
\usepackage{envmath}
\usepackage{subfigure}
\usepackage{graphicx}
\usepackage{natbib}
\usepackage{amsmath}

\newtheorem{The}{Theorem}[section]
\newtheorem{proposition}[The]{Proposition}

\newtheorem{alg}[The]{Algorithm}

\newcommand{\Z}{\bm{Z}}
\newcommand{\T}{\bm{T}}

\begin{document}

\title{Fixed effects Selection in high dimensional Linear Mixed Models} 

\author{Florian Rohart$^{1,2}$, Magali San-Cristobal $^{2}$ and B\'eatrice Laurent $^{1}$\\
\small $^{1}$ UMR 5219, Institut de Math\'ematiques de Toulouse, \\\small INSA de Toulouse, 135 Avenue de Rangueil, 31077 Toulouse cedex 4, France\\\\
\small $^{2}$ UMR 444 Laboratoire de G\'{e}n\'{e}tique Cellulaire, \\\small INRA Toulouse, 31320 Castanet Tolosan cedex, France\\
}

\date{2012}
\maketitle

\begin{abstract}
We consider linear mixed models in which the observations are grouped. A $\ell^1$-penalization on the fixed effects coefficients of the log-likelihood obtained by considering the random effects as missing values is proposed. A multicycle ECM algorithm is used to solve the optimization problem; it can be combined with any variable selection method developed for linear models. The algorithm allows the number of parameters $p$ to be larger than the total number of observations $n$; it is faster than the lmmLasso \citep{Schelldorfer:2011} since no $n\times n$ matrix has to be inverted. We show that the theoretical results of \cite{Schelldorfer:2011} apply for our method when the variances of both the random effects and the residuals are known. The combination of the algorithm with a variable selection method \citep{Rohart:2011} shows good results in estimating the set of relevant fixed effects coefficients as well as estimating the variances; it outperforms the lmmLasso  both in the common case $(p< n)$ and in the high-dimensional case $(p \ge n)$.

\end{abstract}

\section{Introduction}

More and more real data sets are high-dimensional data because of the widely-used new technologies such as high-thoughput DNA/RNA chips or RNA seq in biology. The high-dimensional setting -in which the number of parameters $p$ is greater than the number of observations $n$- generally implies that the problem can not be solved. In order to address this problem, some conditions are usually 
added such as a sparsity condition -which means that a lot of parameters are equal to zero- or a well-conditioning of the variance matrix of the observations, among others. A lot of work has been done to address the problem of variable selection, mainly in a linear model $Y=X\beta+\epsilon$, where $X$ is an $n\times p$ matrix containing the observations and $\epsilon$ is a n-vector of i.i.d random variables, usually Gaussian. One of the oldest method is the Akaike Information Criterion (AIC), which is a penalization of the log-likelihood by a function of the number of parameters included in the model. 
  More recently, the Lasso (Least Absolute Shrinkage and Selection Operator) \citep{Tibshirani:1996} revolutionized the field with both a simple and powerful method: $\ell^1$-penalization of the least squares estimate which exactly shrinks to zero some coefficients. The Lasso has some extensions, a group Lasso \citep{Yuan:2007}, an adaptive Lasso \citep{Huang:2008} and a more stable version known as BoLasso \citep{Bach:2009}, for example. 
 A penalization on the likelihood is not the only way to perform variable selection. Indeed statistical testing has also been used recently \citep{Rohart:2011} and it appears to give good results.
 
 In all methods cited above, the observations are supposed to be independent and identically distributed. When a structure information is available, such as family relationships or common environmental effects, these methods are no longer adapted. In a linear mixed model, the observations are assumed to be clustered, hence the variance-covariance matrix $V$ of the observations is no longer diagonal but could be assumed to be block diagonal in some cases.
A lot of literature about linear mixed models concerns the estimation of the variance components, either with a maximum likelihood estimation (ML) \citep{Henderson:1973,Henderson:1953} or a restricted maximum likelihood estimation (REML) which accounts for the loss in degrees of freedom due to fitting fixed effects \citep{Patterson:1971,Harville:1977,Henderson:1984,Foulley:2002}. However, both methods assume that each fixed effect and each random effect is relevant. This assumption might be wrong  and leads to false estimation of the parameters, especially in a high-dimensional analysis. Contrary to the linear model, there is little literature about selection of fixed effects coefficients in a linear mixed model in a high-dimensional setting. 

Both \citet{Bondell:2010} and \citet{Ibrahim:2011} used a penalized likelihood to perform selection of both the fixed and the random effects. However, their simulation studies were only designed in a low dimensional context.
\citet{Bondell:2010} introduced a constrained EM algorithm to solve the optimization problem, however the algorithm does not really cope with the problem of high dimension. 
 To our knowledge, only \citet{Schelldorfer:2011} studied the topic in a high dimensional setting. Their paper introduced an algorithm based on a $\ell^1$-penalization of the maximum likelihood estimator in order to select the relevant fixed effects coefficients. As highlighted in their paper, their algorithm 
  relies on the inversion of the variance matrix of the observations $V$, which can be time-consuming. Finally, their method depends on a regularization parameter that has to be tuned, as for the original Lasso. As this question remains an open problem, they proposed the use of the Bayesian Information Criterion (BIC) to choose the penalty.\\
All methods are usually considered with one grouping factor -meaning one partition of the observations-, which can be sometimes misappropriate when the observations are divided w.r.t two factors or more; for instance when a family relationship and a common environmental effect are considered.
 
We present in this paper another way to perform selection of the fixed effects in a linear mixed model. We propose to consider the random effects as missing data, as done in \citet{Bondell:2010} or in \citet{Foulley:1997}, and to add a $\ell^1$-penalization on the log-likelihood of the complete data. Our method allows the use of several different grouping factors.
We propose a multicycle ECM algorithm \citep{Foulley:1997,McLachlan:2008,Meng:1993} to solve the optimization problem; this algorithm possesses convergence properties. In addition, we show that the use of BIC in order to tune the regularization parameter as proposed by \citet{Schelldorfer:2011} could sometimes turn out to be misappropriate. \\
We give theoretical results when the variances of the observations are known. Due to the design of the algorithm that is decomposed into steps, the algorithm can be combined with any variable selection method built for linear models. Nevertheless, the performance of the combination strongly depends on the variable selection method that is used.
As there is little literature on the selection of the fixed effects in a high-dimensional linear mixed model, we will mainly compare our results to those of \citet{Schelldorfer:2011}. 

This paper extends the analysis on a real data-set coming from a project in which hundreds of pigs have been studied. The aim is to enlighten relationships between some phenotypes of interest and metabolomic data \citep{Rohart:2012}. Linear mixed models are appropriate since the observations are repeated data from different environments (groups of animals are reared together in the same conditions). Some individuals are also genetically related, in a family effect. 
The data set consists in $506$ individuals from $3$ breeds, $8$ environments and $157$ families. 
The metabolomic data contains $p=375$ variables. We will investigate the Daily Feed Intake (DFI) phenotype.\\


This paper is organized as follows: we will first describe the linear mixed model and the objective function, then we will present the multicycle ECM algorithm that is used to solve the optimization problem of the objective function. Section \ref{improv} gives a generalization of the algorithm of Section \ref{sec1} that can be used with any variable selection method developed for linear models. 
Finally, we will present results from a simulation study showing that the combination of this new algorithm with a good variable selection method performs well, in terms of selection of both the fixed and random effects coefficients (Section \ref{results}), before applying the method on a real data set in Section \ref{real_data}.

\section{The method}
\label{sec1}
Let us introduce some notations that will be used throughout the paper. 
$Var({a})$ denotes the variance-covariance matrix of the vector $ {a}$. For all $a>0$, set $I_a$ to be the identity matrix of $\mathbb{R}^a$. 
For $A\in \mathbb{R}^{n\times p}$, let $A_{I,J}$ $A_{.,J}$ and $A_{I,.}$ denote respectively the submatrix of $A$ composed of elements of $A$ whose rows are in $I$ and columns are in $J$, whose columns are in $J$ with all rows, and whose rows are in $I$ with all columns. Moreover, we set for all $a>0,b>0$,  $0_{a}$ to be the vector of size $a$ with all its coordinates equal to $0$ and $0_{a\times b}$ to be the null matrix of size $a\times b$. Let us denote $|A|$ the determinant of matrix $A$.

\subsection{The linear mixed model setup}
We consider the linear mixed model in which the observations are grouped and we suppose that only a small subset of the fixed effects coefficients are non-zero. The aim of this paper is to recover this subset through an algorithm that will be presented in the next section. 
In the present section we explicit the linear mixed model and our objective function.

Mixed models are often considered with a single grouping factor, meaning that each observation belongs to one single group. In this paper we allow several grouping factors. Assume there are $q$ random effects and $q$ grouping factors $(q\ge1)$, where some grouping factors may be identical. The levels of the factor $k$ are denoted $\{1,2,\dots,N_k\}$. The $i^{th}$-observation belongs to the groups $(i_1,\dots,i_q)$, where for all $l=1,\dots,q$, $i_l \in \{1,2,\dots,N_l\}$. We precise that
two observations can belong to the same group of one grouping factor whereas they can belong to different groups of another grouping factor.

Let $n$ be the total number of observations with $n=\sum_{i=1}^{N_k}n_{i,k}, \forall  k \le q$, where $n_{i,k}$ is the number of observations within group $i$ from the grouping factor $k$. Denote $N=\sum_{k=1}^{q}N_k$.

%
%
The linear mixed model can be written as 
\begin{equation}
\label{mod1}
y=X\beta+\sum_{k=1}^qZ_ku_k+\epsilon,
\end{equation}
where 
\begin{itemize}
\item $y$ is the set of observed data of length $n$,
\item $\beta$ is an unknown vector of $\mathbb{R}^p$; $\beta=(\beta_1,\dots,\beta_p)$,
\item $X$ is the $n\times p$ matrix of fixed effects; $X=(X_1,\dots,X_p)$,
\item For $k=1,\dots,q$, $u_k$ is a $N_k$-vector of the random effect corresponding to the grouping factor $k$, , 
\item  For $k=1,\dots,q$, $Z_k$ is a $n\times N_k$ incidence matrix corresponding to the grouping factor $k$, 
\item $\epsilon=(\epsilon_1,\dots,\epsilon_n)'$ is a Gaussian vector with i.i.d. components $\epsilon\sim \mathcal{N}_n(0,\sigma_e^2I_n)$, where $\sigma_e$ is an unknown positive quantity. We denote by $R$ the variance-covariance matrix of $\epsilon$, $R=\sigma_e^2I_n$.
\end{itemize}
To fix ideas, let us give a example of matrices $Z_k$ for $n=6$ and two random effects.\\ Let $Z_1=\begin{pmatrix}1&0&0\\1&0&0\\0&1&0\\0&1&0\\0&0&1\\0&0&1\end{pmatrix}$ and $Z_2=\begin{pmatrix}x_{1}&0&0\\x_{2}&0&0\\0&x_3&0\\0&x_4&0\\0&0&x_5\\0&0&x_6\end{pmatrix}$. The grouping factors 1 and 2 are the same for the two random effects $u_1$ and $u_2$, and  $Z_2$ is the incidence matrix of the interaction of the variable $x=(x_1,\dots,x_6)$ and the grouping factor.

Throughout the paper, we assume that $u_k\sim \mathcal{N}_{N_k}(0,\sigma_k^2I_{N_k})$, where $\sigma_k$ is an unknown positive quantity. 
We denote $u=(u_1',\dots,u_k')'$, $Z$ the concatenation of $(Z_1,\dots,Z_q)$, $G$ the block diagonal matrix of $\sigma_1^2I_{N_1},\dots,\sigma_q^2I_{N_q}$ and $\Gamma$ the block diagonal matrix of $\gamma_1I_{N_1},\dots,\gamma_qI_{N_q}$, where  $\gamma_k=\sigma_e^{2}/\sigma_k^{2}$.\\
Remark that with these notations, Model \eqref{mod1} can also be written as: $y=X\beta+Zu+\epsilon$.\\
In the following, we assume that $\epsilon, {u}_1,\dots, {u}_q$ are mutually independent. Thus $Var( {u}_1,\dots, {u}_q,\epsilon)=\begin{pmatrix}G& {0}\\ {0}&R\end{pmatrix}$. 
We consider the matrices $X$ and $\{Z_k\}_{1,\dots,q}$ to be fixed design.

Note that our model \eqref{mod1} and the one in \cite{Schelldorfer:2011} are almost identical when all the grouping factors are identical, except that we supposed $u_1\dots,u_q$ to be independent while they did not make this assumption. Nevertheless, for their simulation study, they considered  i.i.d. random effects.\\

Let us denote by $J$ the set of the indices of the relevant fixed effects of Model \eqref{mod1}; $J=\{j,\beta_j\neq 0\}$. The aim of this paper is to estimate $J$, $\beta$, $G$ and $R$.
In the whole paper, the number of fixed effects $p$ can be larger than the total number of observations $n$. However, we focus on the case where only a few fixed-effects are relevant. We also assume that only a few grouping factors are included in the model since this paper was motivated by such a case on a real data set, see Section \ref{real_data}. Hence we assume $N+|J|<n$.  

\subsection{A $\ell^1$ penalization of the complete log-likelihood}


In the following, we consider the fixed effects coefficients $\beta$ and the variances $\sigma_1^2,\dots,\sigma_q^2,\sigma_e^2$ as parameters and $\{u_k\}_{k\in\{1,\dots,q\}}$ as missing data. 
 We denote $\Phi=(\beta,\sigma_1^2,\dots,\sigma_q^2,\sigma_e^2)$.
 \\ The log-likelihood of the complete data $ {x}=( {y}, {u})$ is 
\begin{equation}
\label{loglik}
L(\Phi; {x})=L_0(\beta,\sigma_e^2,\sigma_1^2,\dots,\sigma_q^2;\epsilon)+\sum_{k=1}^qL_k(\sigma_k^2;u_k),
\end{equation}
where 
\begin{subequations}
\begin{gather}
\label{L0}
-2L_0(\beta,\sigma_e^2,\sigma_1^2,\dots,\sigma_q^2;\epsilon)=n \log(2\pi)+n \log(\sigma_e^2) +\left|\left|y-X\beta-\sum_{k=1}^qZ_ku_k\right|\right|^2/\sigma_e^2,\\
\label{L}
\forall k\in \{1,\dots,q\}, -2L_k(\sigma_k^2;u_k)=N_k \log(2\pi)+N_k \log(\sigma_k^2)+\left|\left|u_k\right|\right|^2/\sigma_k^2.
\end{gather}
\end{subequations}

Indeed, \eqref{loglik} comes from $p(x|\Phi)=p(y|\beta,u_1,\dots,u_q,\sigma_e^2)\Pi_{k=1}^q p(u|\sigma_k^2)$; \eqref{L0} comes from $L_0(\beta,\sigma_e^2,\sigma_1^2,\dots,\sigma_q^2;\epsilon)=L_0(\sigma_e^2;\epsilon)=n\log(2\pi)+n\log(\sigma_e^2) +\epsilon'\epsilon/\sigma_e^2$ because $\epsilon|\sigma_e^2\sim\mathcal{N}_n(0,\sigma_e^2I_n)$ and \eqref{L} from $u_k|\sigma_k^2\sim\mathcal{N}_{N_k}(0,\sigma_k^2I_{N_k})$.

Since we allow the number of fixed-effects $p$ to be larger than the total number of observations $n$, the usual maximum likelihood  (ML) or restricted maximum likelihood (REML) approaches do not apply. As we assumed that $\beta$ is sparse  -many coefficients are assumed to be null- and since we want to recover that sparsity, 
we add a $\ell^1$ penalty on $\beta$ to the log-likelihood of the complete data \eqref{loglik}. Indeed a $\ell^1$ penalization is known to induce sparsity in the solution, as in the Lasso method \citep{Tibshirani:1996} or the lmmLasso method \citep{Schelldorfer:2011}. Thus we consider the following objective function to be minimized:
 \begin{equation}
\label{gtheta}
 g(\Phi; {x})=-2L(\Phi; {x})+\lambda|\beta|_1,
 \end{equation}
 where $\lambda$ is a positive regularization parameter. Remark that the function $g$ could have been obtained from a Bayesian setting considering a Laplace prior on $\beta$.\\
It is interesting to note that finding a minimum of the objective function $\eqref{gtheta}$ is a non-linear, non-differentiable and non convex problem. But more importantly, one thing that strikes out -especially from \eqref{L}- is that the function $g$ is not lower-bounded. Indeed, $L(\Phi; {x})$ tends to infinity when both $u_k$ and $\sigma_k$ tends toward $0$. 
It is a well-known problem of degeneracy of the likelihood, especially studied in Gaussian mixture model \citep{Biernacki:2003} but not much concerning mixed models. In linear mixed models, some authors focus on the log-likelihood of the marginal model in which the random effects are integrated out in the matrix of variance of the observations $Y$, such as in \citet{Schelldorfer:2011}:
\begin{equation*}
y=X\beta+\epsilon,\text{ where } \epsilon\sim\mathcal{N}(0,V).
\end{equation*}
 Note that $V=ZGZ'+R$. The degeneracy of the likelihood can also appear in the marginal model when the determinant of $V$ tends toward zero. This phenomenon is likely to happen  in a high dimensional context when too much fixed-effects enter the model, that is to say when the amount of regularization chosen by the penalty of the lmmLasso \citep{Schelldorfer:2011} or by $\lambda$ in \eqref{gtheta}  is not large enough.  

Because of the non lower-boundness of the likelihood, the problem of minimizing the function $g$ is ill-posed: we are not interested in the minimization of $g$ on the parameter space $\{\beta\in\mathbb{R}^p,\sigma_1^2\ge0,\dots,\sigma_q^2\ge0,\sigma_e^2\ge0\}$ but more interested in  minimizing $g$ inside the parameter space 
\begin{equation*}
\Lambda=\{\beta\in\mathbb{R}^p,\sigma_1^2>0,\dots,\sigma_q^2>0,\sigma_e^2>0\}.
\end{equation*}

Instead of adding a $\ell^1$ penalty on the random effect as \cite{Bondell:2010}, 
 we will use the degeneracy of the likelihood at the frontier of the parameter space $\Lambda$ to perform selection of the random effects. Indeed, if it exists $1\le k\le q$ such that the minimization process of the function $g$, defined by \eqref{gtheta}, takes place at the frontier $\sigma_k^2=0$ of the parameter space $\Lambda$, then the grouping factor $k$ is deleted from the model \eqref{mod1}.
 Nevertheless, our method is more restrictive than the one of \cite{Bondell:2010} since we assume $N+|J|<n$. \\
 
The minimization process of the function $g$ can coincide with the deletion of the random effect $k$, for $1\le k \le q$, for two reasons: either the true underlying model was different from the fitted one -some grouping factors are included in the model although there is no need to-, or because the initialization of the minimization process was to close to an attraction domain of $(u_k,\sigma_k^2)=(0_{N_k},0)$ \citep{Biernacki:2003}. 

When selection of the random effects is performed in the linear mixed model \eqref{mod1} with $q$ random effects, a new model is fitted with $q-1$ grouping factor and the objective function is modified accordingly. The selection of the random effects can be performed until no grouping factor remains, then a linear model is considered. 

 In the next section we will use a multicycle ECM algorithm in order to solve the minimization of \eqref{gtheta}; it performs selection of both the fixed and the random effects.

\subsection{A multicycle ECM algorithm}
\label{alg}

The multicycle ECM algorithm \citep{Meng:1993,Foulley:1997,McLachlan:2008} used to solve the minimization problem of \eqref{gtheta} contains four steps -two E steps interlaced with two M steps-; each will be described in this section. \\
Recall that $\Phi=(\beta,\sigma_1^2,\dots,\sigma_q^2,\sigma_e^2)$ is the vector of the parameters to estimate and that $u=(u_1',\dots,u_k')'$ is a vector of missing values. 
For the sake of simplicity, we denote $\mathcal{K}=\{1,\dots,q\}$ and $\sigma_{\mathcal{K}}^2=\{\sigma_k^2\}_{k\in \mathcal{K}}$.\\
%
%
%
 The multicyle ECM algorithm is an iterative algorithm. We will index the iterations by $t\in\mathbb{N}$. $\Theta^{[t]}$ will denote the current estimation of the parameter $\Theta$ at iteration $t$.
 
Let $E_{u|y,\Phi=\Phi^{[t]}}$ denote the conditional expectation under the distribution of $u$ given the vector of observations $y$ and the current estimation of the set of parameters $\Phi$ at iteration $t$.

\subsubsection{First E-step}

Let denote 
\begin{equation*}
Q(\Phi;\Phi^{[t]})=E_{u|y,\Phi=\Phi^{[t]}}[g(\Phi;  x)].
\end{equation*}
We can decompose $Q$ as follows:\\
 \begin{equation*} 
Q(\Phi;\Phi^{[t]})=Q_0(\beta,\sigma_{\mathcal{K}}^2,\sigma_e^2;\Phi^{[t]})+\sum_{k=1}^qQ_k(\sigma_k^2;\Phi^{[t]}),
\end{equation*} 
where
\begin{equation*}
Q_0(\Phi;\Phi^{[t]})= n\ \log(2\pi)+n\ \log(\sigma_e^{2[t]}) +E_{u|  y,\Phi=\Phi^{[t]}}(\epsilon'\epsilon)/\sigma_e^{2[t]}+\lambda|\beta^{[t]}|_1\\
\end{equation*}
and\begin{equation*}
\forall k\in {\mathcal{K}}, Q_k(\sigma_k^{2};\Phi^{[t]})=N\log(2\pi)+N\log(\sigma_k^{2[t]})+E_{u|  y,\Phi=\Phi^{[t]}}({u_k}'u_k)/\sigma_k^{2[t]}.
\end{equation*}
By definition, we have for all $1\le i \le n, Var_{u|  y,\Phi=\Phi^{(t)}}\left(\epsilon_i\right)= E_{u|  y,\Phi=\Phi^{[t]}}(\epsilon_i^2)-\left|\left|E_{u|  y,\Phi=\Phi^{[t]}}\left(\epsilon_i\right)\right|\right|^2$.\\ Hence
\begin{equation*}
E_{u|  y,\Phi=\Phi^{[t]}}(\epsilon'\epsilon)=\left|\left|E_{u|  y,\Phi=\Phi^{[t]}}\left(\epsilon\right)\right|\right|^2+tr\left(Var_{u|  y,\Phi=\Phi^{(t)}}\left(\epsilon\right)\right).
\end{equation*}
We can then explicit 
\begin{equation}
\label{u1}
E_{u|y,\Phi=\Phi^{[t]}}(\epsilon'\epsilon)=\left|\left|   y-X\beta^{[t]}-Z E\left(u|  y,\Phi=\Phi^{[t]}\right)\right|\right|^2+tr\left(Z Var\left(u|  y,\Phi^{[t]}\right)Z'\right).
\end{equation}
%
According to the denomination of \citet{Henderson:1973}, $E\left(u|y,\Phi=\Phi^{[t]}\right)$ is the BLUP (Best Linear Unbiased Prediction) of $u$ for the vector of parameters $\Phi$ equal to $\Phi^{[t]}$. Let us denote $u^{[t+1/2]}=E\left(u|y,\Phi=\Phi^{[t]}\right)$, we have that
\begin{equation*}
 u^{[t+1/2]}=(Z'Z +\Gamma^{[t]})^{-1}Z'\left(  y-X\beta^{[t]}\right).
\end{equation*}


\subsubsection{M-Step for $\beta$}
The next step performs a minimization of $Q_0(\beta,\sigma_\mathcal{K}^2,\sigma_e^2;\Phi^{[t]})$ with respect to $\beta$:
\begin{equation}
\label{Lasso}
\beta^{[t+1]}=\underset{\beta}{Argmin} \left( \dfrac{1}{\sigma_e^{2[t]}}\left|\left|\left(  y-Zu^{[t+1/2]}\right)-X\beta\right|\right|^2+\lambda\left|\beta\right|_1\right).
\end{equation}
Remark that \eqref{Lasso} is a Lasso on $\beta$ with the vector of ``observed" data $\left(  y-Zu^{[t+1/2]}\right)$ and the penalty $\lambda\sigma_e^{2[t]}$. \\

\subsubsection{Second E-Step}
A second E-step is performed with the actualization of the vector of missing values $u$: $u^{[t+1]}=E\left(u|y,\beta=\beta^{[t+1]},\sigma_1^2=\sigma_1^{2[t]},\dots,\sigma_q^2=\sigma_q^{2[t]},\sigma_e^2=\sigma_e^{2[t]} \right)$, thus
\begin{equation*}
\label{ridgek}
u^{[t+1]}=(Z'Z +\Gamma^{[t]})^{-1}Z'\left(  y-X\beta^{[t+1]}\right).
\end{equation*}
We define $\forall k \in \mathcal{K}$, $u_k^{[t+1]}$ to be the element of size $N_k$ that corresponds to the grouping factor $k$ in $u^{[t+1]}$.\\

\subsubsection{M-step for $(\sigma_1^2,\dots,\sigma_q^2,\sigma_e^2)$}
The actualization of the variances $\{\sigma_k^2\}_{1\le k\le q}$ and $\sigma_e^2$ are performed with the minimization of $\{Q_k\}_{1\le k\le q}$ and $Q_0$ respectively.\\
 Let $k\in \mathcal{K}$, the minimization of $Q_k$ with respect to $\sigma_k^2$ gives:\\
$\sigma_k^{2[t+1]}=E\left(u_k'u_k|  y,\sigma_k^{2[t]},\sigma_e^{2[t]},\beta^{[t+1]}\right)/N_k$.\\
Besides, 
\begin{equation*}
E\left(u_k'u_k|  y,\sigma_k^{2[t]},\sigma_e^{2[t]},\beta^{[t+1]}\right)=\left|\left|E\left(u_k|  y,\sigma_k^{2[t]},\sigma_e^{2[t]},\beta^{[t+1]}\right)\right|\right|^2+tr\left(Var\left(u_k|  y,\sigma_k^{2[t]},\sigma_e^{2[t]},\beta^{[t+1]}\right)\right).
\end{equation*}
Moreover we have, thanks to \cite{Henderson:1973},
\begin{equation*}
Var\left(u_k|  y,\sigma_k^{2[t]},\sigma_e^{2[t]},\beta^{[t+1]}\right)=T_{k,k}\sigma_e^{2[t]},
\end{equation*}
where $T_{k,k}$ is defined as follows:
\begin{eqnarray*}
\left(Z'Z +\Gamma^{[t]}\right)^{-1} &=&\begin{pmatrix}Z_1'Z_1+\gamma_1^{[t]} I_{N_1} & Z_1'Z_2&  \hdots &Z_1'Z_q\\Z_2'Z_1 &Z_2'Z_2+\gamma_2^{[t]} I_{N_2} & \hdots & Z_2'Z_q \\\vdots & \vdots &\ddots &\vdots\\ Z_q'Z_1 &Z_q'Z_2& \hdots&Z_q'Z_q+\gamma_q^{[t]} I_{N_q}\end{pmatrix}^{-1}\\
&=&\begin{pmatrix}T_{1,1} & T_{1,2}& \hdots & T_{1,q}\\T_{1,2}'& T_{2,2}&\hdots& T_{2,q}\\\vdots &\vdots& \ddots &\vdots\\T_{1,q}'& T_{2,q}'&\hdots & T_{q,q}\end{pmatrix}.
\end{eqnarray*}
Thus, for all $k \in \mathcal{K}$:
\begin{equation*}
\label{sigmau}
\sigma_k^{2[t+1]}=\dfrac{1}{N_k}\left[\left|\left| u_k^{[t+1]}\right|\right|^2+tr\left(T_{k,k}\right)\sigma_e^{2[t]}\right].
\end{equation*}
The minimization of $Q_0$ with respect to $\sigma_e^2$ gives:
$\sigma_e^{2[t+1]}=E_{u|  y,\Phi=\Phi^{[t]}}(\epsilon'\epsilon)/n$. From \eqref{u1}, we have
\begin{equation*}
\label{sigmae1}
\sigma_e^{2[t+1]}=\dfrac{1}{n}\left[\left|\left|  y-X\beta^{[t+1]}-Z u^{[t+1]}\right|\right|^2+tr\left(Z(Z'Z +\Gamma^{[t]})^{-1}Z'\right)\sigma_e^{2[t]}\right].
\end{equation*}
Since
\begin{eqnarray*}
tr\left(Z\left(Z'Z +\Gamma^{(t)}\right)^{-1}Z'\right)&=&tr\left(\left(Z'Z +\Gamma^{(t)}\right)^{-1}Z'Z\right)\\
&=& N-tr\left[\left(Z'Z +\Gamma^{(t)}\right)^{-1}\Gamma^{[t]}\right]\\
&=&N-\sum_{k=1}^q \gamma_k^{[t]}tr\left(T_{k,k}\right)
\end{eqnarray*}
we have
\begin{equation*}
\sigma_e^{2[t+1]}=\dfrac{1}{n}\left[\left|\left|  y-X\beta^{[t+1]}-Z u^{[t+1]}\right|\right|^2+\left(N-\sum_{k=1}^q\gamma_k^{[t]}tr\left(T_{k,k}\right)\right)\sigma_e^{2[t]}\right].
\end{equation*}
In summary, the algorithm is the following:
\begin{alg}[Lasso+]
\label{alg1}
\begin{flushleft}
\underline{Initialization:}\\
Set $\mathcal{K}=\{1,\dots,q\}$. Initialize the set of parameters $\Phi^{[0]}=(\sigma_{\mathcal{K}}^{2[0]},\sigma_e^{2[0]},\beta^{[0]})$. \\
Define $\Gamma^{[0]}$ as the block diagonal matrix of $\gamma_1^{[0]}I_{N_1},\dots,\gamma_q^{[0]}I_{N_q}$, where $\gamma_k^{[0]}=\sigma_e^{2[0]}/\sigma_k^{2[0]}$. Define $Z$  as the concatenation of $Z_1,\dots,Z_q$ and $u=(u_1',\dots,u_q')'$.\\
\underline{Until convergence:}\\
 1. \textit{E-step}\\
 $u^{[t+1/2]}=(Z'Z +\Gamma^{[t]})^{-1}Z'(y-X\beta^{[t]})$\\
 2. \textit{M-step}\\
$\beta^{[t+1]}=\underset{\beta}{Argmin} \left( \left|\left|\left(y-Z u^{[t+1/2]}\right)-X\beta\right|\right|^2+\lambda\sigma_e^{2[t]}\left|\beta\right|_1\right)$\\
 3. \textit{E-step}\\
 $u^{[t+1]}=(Z'Z +\Gamma^{[t]})^{-1}Z'(y-X\beta^{[t+1]})$\\
 4. \textit{M-step}\\
 (a) For k in $\mathcal{K}$, 
set ${\sigma^2_{k}}^{[t+1]}=\left|\left| u^{[t+1]}_{k}\right|\right|^2/N_k+tr\left(T_{k,k}\right)\sigma_e^{2[t]}/N_k$ \\
(b) Set $\sigma_e^{2[t+1]}=\dfrac{1}{n}\left[\left|\left|y-X\beta^{[t+1]}-Z u^{[t+1]}\right|\right|^2+\sum_{k\in\mathcal{K}}\left(N_k-\gamma_k^{[t]}tr\left(T_{k,k}\right)\right)\sigma_e^{2[t]}\right]$\\
(c) For k in $\mathcal{K}$, if $\left(\left|\left| u^{[t+1]}_{k}\right|\right|^2/N_k<10^{-4}\sigma_e^{2[t]}\right)$ then $\mathcal{K}=\mathcal{K}\backslash \{k\}$\\
Define $Z$  as the concatenation of $\{Z_k\}_{k\in\mathcal{K}}$ and $u$ as the transpose of the concatenation of $\{u_k'\}_{k\in\mathcal{K}}$.\\
Set $\Gamma^{[t+1]}$ as the block diagonal matrix of $\{\gamma_k^{[t+1]}I_{N_k}\}_{k\in\mathcal{K}}$, 
where for all $k\in\mathcal{K}$, $\gamma_k^{[t+1]}=\sigma_e^{2[t+1]}/\sigma_k^{2[t+1]}$. \\
\underline{end}
\end{flushleft}
\end{alg}

The convergence of Algorithm \ref{alg1} is ensured since it is a multicycle ECM algorithm \citep{Meng:1993}.
\\
Three stopping criteria are used to stop the convergence process of the algorithm: a condition on $||\beta^{[t+1]}-\beta^{[t]}||^2$, a condition on $||u_k^{[t+1]}-u_k^{[t]}||^2$ for each random effect $u_k$ and a condition on $||L(\Phi^{[t+1]},x)-L(\Phi^{[t]},x)||^2$ where $L(\Phi,x)$ is the log-likelihood defined by \eqref{loglik}. The convergence takes place when all the criteria are fulfilled. 
We also add a fourth condition that controls the number of iterations.
We choose to initialize the algorithm \ref{alg1} as follows: 
for all $1\le k\le q, \sigma_{k}^{2[0]}=\frac{0.4}{q} \ \sigma_e^{2[-1]}, \sigma_e^{2[0]}=0.6\ \sigma_e^{2[-1]}$, and $(\sigma_e^{2[-1]},\beta^{[0]})$ is estimated from a linear estimation (without the random effects) of the Lasso at the given penalty $\lambda$. We will study in Section \ref{influence} the influence of the initialization of the algorithm on simulated data.

Note that Step 4(c) performs the selection on the random effects; we decide to delete a random effect when its variance became lower that $10^{-4}\sigma_e^{2[t]}$.\\

\label{bias}
The estimation of the set of parameters $\Phi$ is biased \citep{Zhang:2008}. One last step can be added in order to address this problem once both Algorithm \ref{alg1} has converged and the penalization parameter $\lambda$ has be tuned. Indeed, one should prefer to use Algorithm \ref{alg1} in order to estimate both the support of $\beta$ and the support of the random effects, and then to estimate the set $\Phi$ with a classical mixed model estimation on the model:
\begin{equation*}
\label{modbias}
y=X\beta_{\hat J}+\sum_{k\in S}Z_ku_k+\epsilon,
\end{equation*}
where $\hat J$ and $S$ are the estimated set of indices of the relevant fixed effects and the estimated set of indices of the relevant random effects respectively.

\begin{proposition}
\label{prop}
When the variances are known, the minimization of our objective function \eqref{gtheta} is the same as the minimization of  $Q(\beta)=(y-X\beta)'V^{-1}(y-X\beta)+\lambda|\beta|_1$, which is the objective function of \cite{Schelldorfer:2011} at known variances.
\end{proposition}
Let us recall that \cite{Schelldorfer:2011} obtained theoretical results on the consistency of their method. According to Proposition \ref{prop}, these results apply to our method in the case of known variances.
The proof of Proposition \ref{prop} is given in Appendix C.\\

Note that when individuals are genetically related through a known relationship matrix $A$, we have $u\sim \mathcal{N}_n(0,\sigma_s^2A)$, with $\sigma_s>0$. Thanks to \cite{Henderson:1973}, $A^{-1}$ can be directly computed. In all that precede, the changes are the following : the matrix $\Gamma$ becomes the matrix $\sigma_e^2/\sigma_s^2A^{-1}$ and $||u||^2$ becomes $u'A^{-1}u$. \\

\subsection{The tuning parameter}
\label{tun}

Algorithms \ref{alg1} involves a regularization parameter $\lambda$; the solution depends on this parameter. This amount of shrinkage  has to be tuned. We choose the use of the Bayesian Information Criterion (BIC) \citep{Schwarz:78}:
\begin{equation*}
\label{BIC}
\lambda_{BIC}=\underset{\lambda}{Argmin}\{ \log|V_{\lambda}|+(y-X\hat\beta_{\lambda})'V_{\lambda}^{-1}(y-X\hat\beta_{\lambda})+d_\lambda. \log(n)\},
\end{equation*}
where $V_\lambda=\sum_{k\in \mathcal{K}}\hat \sigma_k^2Z_k Z'_k+ \hat \sigma_e^2I_n$ and $\hat\sigma_k^2, \hat\sigma_e^2,\hat\beta_\lambda$ are obtained from the minimization of the objective function $g$ defined by \eqref{gtheta}.  
Moreover, $d_\lambda:=\sum_{k=1}^p1_{\sigma_k\neq 0}+|\hat J_{\lambda}|$ is the sum of the number of non-zero variance-covariance parameters and the number of non-zero fixed effects coefficients included in the model which has been selected with the regularization parameter $\lambda$.\

 Other methods can be used to choose $\lambda$ such as AIC or cross-validation, among others. An advantage of BIC over cross-validation is mainly the gain of computational time. 
 
 In the next section, we propose a generalization of Algorithm \ref{alg1} which allows the use of any variable selection methods developed for linear models.

\section{A generalized algorithm}
\label{improv}
Algorithm \ref{alg1} gives good results, as it can be seen in the simulation study of Section \ref{results}. Nevertheless, since Step 2 of Algorithm \ref{alg1} aims at selecting the relevant coefficients of $\beta$ in a linear model, the Lasso method can be replaced with any variable selection method built for linear models. 
If the chosen variable selection method optimizes a criterion, such as the adaptive Lasso \citep{Zou:2006} or the elastic net \citep{Zou:2005}, the algorithm thus obtained remains a multicycle ECM algorithm and the convergence property still applies. However, the convergence property does not hold for methods that do not optimize a criterion.

Algorithm \ref{alg1} can be reshaped for a generalized algorithm as follows:
\begin{alg}
\label{alg3}
\begin{flushleft}
\underline{Initialization:}\\
Initialize the set of parameters $\Phi^{[0]}=(\sigma_{\mathcal{K}}^{2[0]},\sigma_e^{2[0]},\beta^{[0]})$. Set $\mathcal{K}=\{1,\dots,q\}$.\\
Define $\Gamma^{[0]}$ as the block diagonal matrix of $\gamma_1^{[0]}I_{N_1},\dots,\gamma_q^{[0]}I_{N_q}$, where $\gamma_k^{[0]}=\sigma_e^{2[0]}/\sigma_k^{2[0]}$. Define $Z$  as the concatenation of $Z_1,\dots,Z_q$ and $u=(u_1',\dots,u_q')'$.\\
\underline{Until convergence:}\\
 1.  $u^{[t+1/2]}=(Z'Z +\Gamma^{[t]})^{-1}Z'(y-X\beta^{[t]})$\\
 2.  Variable selection and estimation of $\beta$ in the linear model  $y-Zu^{[t+1/2]}=X\beta+\epsilon^{[t]}$,  where $\epsilon^{[t]}\sim\mathcal{N}(0,\sigma_e^{2[t]}I_n)$.\\
 3. $u^{[t+1]}=(Z'Z +\Gamma^{[t]})^{-1}Z'(y-X\beta^{[t+1]})$\\
4. 
 (a) For k in $\mathcal{K}$, 
set ${\sigma^2_{k}}^{[t+1]}=\left|\left| u^{[t+1]}_{k}\right|\right|^2/N_k+tr\left(T_{k,k}\right)\sigma_e^{2[t]}/N_k$ \\
(b) Set $\sigma_e^{2(t+1)}=\dfrac{1}{n}\left[\left|\left|y-X\beta^{[t+1]}-Z u^{[t+1]}\right|\right|^2+\sum_{k\in\mathcal{K}}\left(N_k-\gamma_k^{[t]}tr\left(T_{k,k}\right)\right)\sigma_e^{2[t]}\right]$\\
(c) For k in $\mathcal{K}$, if $\left(\left|\left| u^{[t+1]}_{k}\right|\right|^2/N_k<10^{-4}\sigma_e^{2[t]}\right)$ then $\mathcal{K}=\mathcal{K}\backslash \{k\}$\\
Define $Z$  as the concatenation of $\{Z_k\}_{k\in\mathcal{K}}$ and $u$ as the transpose of the concatenation of $\{u_k'\}_{k\in\mathcal{K}}$.\\
Set $\Gamma^{[t+1]}$ as the block diagonal matrix of $\{\gamma_k^{[t+1]}I_{N_k}\}_{k\in\mathcal{K}}$, 
where for all $k\in\mathcal{K}$, $\gamma_k^{[t+1]}=\sigma_e^{2[t+1]}/\sigma_k^{2[t+1]}$. \\
\underline{end}
\end{flushleft}
\end{alg}

We choose to initialize Algorithm \ref{alg3} as follows: 
for all $1\le k\le q, \sigma_{k}^{2[0]}=\frac{0.4}{q} \ \sigma_e^{2[-1]}, \sigma_e^{2[0]}=0.6\ \sigma_e^{2[-1]}$, and $(\sigma_e^{2[-1]},\beta^{[0]})$ is estimated from a linear estimation (without the random effects) of the method used at Step 2. \\
In the following we propose to combine Algorithm \ref{alg1} with a method that does not need a tuning parameter, namely the procbol method \citep{Rohart:2011}. The procbol method is a sequential multiple hypotheses testing which statistically determines the set of relevant variables in a linear model $y=X\beta +\epsilon$ where $\epsilon$ is an i.i.d Gaussian noise. This method is a two-step procedure: the first step orders the variables taking into account the observations $y$ and the second step uses multiple hypotheses testing to separate the relevant variables from the irrelevant ones. 
The procbol method is proved to be powerful under some conditions on the signal in \cite{Rohart:2011}.

In Section \ref{results}, we show that the combination of Algorithm \ref{alg3} and the procbol method performs well on simulated data. 

%

\section{Simulation study}
\label{results}
The purpose of this section is to compare different methods that aim at selecting both the correct fixed effects coefficients and the relevant random effects in a linear mixed model \eqref{mod1}, but also to look at the improvement obtained from including random effects in the model. 
\subsection{Presentation of the methods}
We compare several methods, some of them are designed to work in a linear model: \textit{Lasso} \citep{Tibshirani:1996}, \textit{adLasso} \citep{Zou:2006} and \textit{procbol} \citep{Rohart:2011}, while others are designed to work in a linear mixed model: \textit{lmmLasso} \citep{Schelldorfer:2011}, \textit{Algorithm \ref{alg1}} (labelled as \textit{Lasso+}), \textit{adLasso+Algorithm \ref{alg3}} (labelled as \textit{adLasso+}) and \textit{procbol+Algorithm \ref{alg3}} (labelled as \textit{pbol+}). \\
The initial weights of the \textit{adLasso} and \textit{adLasso+} are set to be equal to $1/|\hat \beta_{i}|$ where for all $i\in\{1,\dots,p\}$, $\hat\beta_i$ is the Ordinary Least Squares (OLS) estimate of $\beta_i$ in the model $y=X_i\beta_i+\epsilon_i$.

The second step of the procbol method performs multiple hypotheses testing with an estimation of unknown quantiles related to the matrix $X$. The calculation of these quantiles at each iteration of the convergence process would make the combination of the procbol method and Algorithm \ref{alg3} almost impossible to run; however, since the data matrix $X$ stays the same throughout the algorithm, the quantiles also do. Thus the procbol method was adapted to be run several times on the same data set by keeping the calculated quantiles, which led to a enormous gain of computational time. Some parameters of the procbol method were changed in order to limit the time of one iteration of the convergence process, as follows. The parameter $m$ which stands  for the number of bootstrapped samples used to sort the variables (first step of the procbol method) was set to $10$. The number of variables ordered at the first step of the procbol method was set to $40$. Note that when the procbol method was used in a linear model, we set $m=100$ as advised in \cite{Rohart:2011}. Both the \textit{procbol} method and the \textit{pbol+} method were set with a user-level of $\alpha\in \{0.1,0.05\}$, which stands for the level of the testing procedure.\\
Concerning all methods that needed a tuning parameter, we set it using the Bayesian Information Criterion described in Section \ref{tun}. 
A particular attention has to be drawn on the tuning of the regularization parameter of some methods that could be tricky in some cases due to the degeneracy of the likelihood, especially \textit{Lasso} and \textit{adLasso}, see Appendix B.

\subsection{Design of our simulation study}
Concerning the design of our simulations, we set $X_1$ to be the vector of $\mathbb{R}^n$ whose coordinates are all equal to $1$ and we considered four models. For each model, the response variable $y$ is computed via $y=\sum_{j=1}^5X_{i_j}\beta_{i_j}+\sum_{k=1}^qZ_ku_k+ \epsilon$, where $J=\{i_1,\dots,i_5\}\subset \{1,\dots,p\}$, with two random effects ($q=2$) being standard Gaussian ($\sigma_1^2=\sigma_2^2=1$) and $\epsilon$ being a vector of independent standard Gaussian variables. The models used to fit the data differ in the number of parameters $p$, the number of random effects $q$ and the dependence structure of the $X_i$'s. 
For each model, we have that for all $j=2,\dots,p$: $\sum_{i=1}^n X_{j,i}=0$ and $\frac{1}{n}\sum_{i=1}^n X_{j,i}^2=1$.
 For $k=1,\dots,q$, the random effects regression matrix $Z_k$ corresponds to the design matrix of the interaction between the $k^{th}$ column of $X$ and the grouping factor $k$, which gives a $n\times N_k$ matrix. 
The design of the matrices $Z_k$'s means that the first $q$ grouping variables generates both a fixed effect (corresponds to $\beta_k$'s) and a random effect (corresponds to $u_k$'s). 
As advised in \citet{Schelldorfer:2011}, the variables that generate both a fixed and a random effect do not undergo feature selection; otherwise the fixed effect coefficients of those variables tends to be shrunken towards 0. The set of variables that do not undergo feature selection can change at each step of the convergence process of our algorithms. Indeed, as soon as a variable does not generate a random effect anymore, the fixed effect corresponding to that variable undergoes feature selection again.\\
The models are defined as follows:
\begin{itemize}
\item $M_1$: $n=120$, $p=80$, $\beta_J=2/3$. For all $j=2,\dots,p, X_j\sim \mathcal{N}_n(0,I_n)$.
The division of the observations for the two random effects are the same; for all $k\le 2: N_k=20,\forall i\in \{1,..,20\} \ n_{i,k}=6$. This model is fitted assuming $q=3$.
\item $M_2$: $n=120$, $p=300$, $\beta_J=3/4$. The covariates are generated from a multivariate normal distribution with mean zero and covariance matrix $\Sigma$ with the pairwise correlation $\Sigma_{kk'}= \rho^{|k-k'|}$ and $\rho=0.5$.
The division of the observations for the two random effects are the same; for all $k\le 2: N_k=20,\forall i\in \{1,..,20\} \ n_{i,k}=6$.
\item $M_3$: $n=120$, $p=300$, $\beta_J=2/3$. For all $j=2,\dots,p, X_j\sim \mathcal{N}_n(0,I_n)$. The division of the observations for the two random effects are different: 
 $N_1=20,\forall i\in \{1,..,20\} \ n_{i,1}=6$ and  $N_2=15, \forall i\in \{1,..,15\} \ n_{i,2}=8$
\item $M_4$: $n=120$, $p=600$, $\beta_J=2/3$. For all $j=2,\dots,p, X_j\sim \mathcal{N}_n(0,I_n)$. The division of the observations for the two random effects are the same; for all $k\le 2: N_k=20,\forall i\in \{1,..,20\} \ n_{i,k}=6$. 
 \end{itemize}
 For models $M_1,M_3,M_4$, we set $J=\{1,\dots,5\}$. For model $M_2$, we set $J=\{1,2,i_3,i_4,i_5\}$ where $\{i_3,i_4,i_5\}\subset\{3,\dots,p\}$.\\

In each model, the aim is to recover both the set of relevant fixed effects coefficients $J$ and the set of relevant random effects; but also to estimate the variance of both the random effects and the residuals. To judge the quality of the methods, we use several criterion: the percentage of true model recovered under the label `Truth' (both $J$ and the set of relevant random effects),  the percentage of times the true set of fixed effects is recovered `$\hat J=J$', 
the cardinal of the estimated set of fixed effects coefficients $|\hat J|$, the number of true positive $TP$, the estimated variance $\hat\sigma_e^2$ of the residuals, the estimated variances $\hat\sigma_1^2,\dots,\hat\sigma_q^2$  of the random effects  and 
the mean squared error $mse$ calculated as an $\ell^2$ error rate between the reality -$X\beta$- and the estimation -$X\hat\beta$-. 
 We also calculated the Signal-to-Noise Ratio (SNR) as $||X\beta||_2^2/||\sum_{k=1}^qZ_ku_k+\epsilon||_2^2$ for each of the replications.

\subsection{Comments on the results}

\begin{figure}
\centering
{\includegraphics[width=7.5cm]{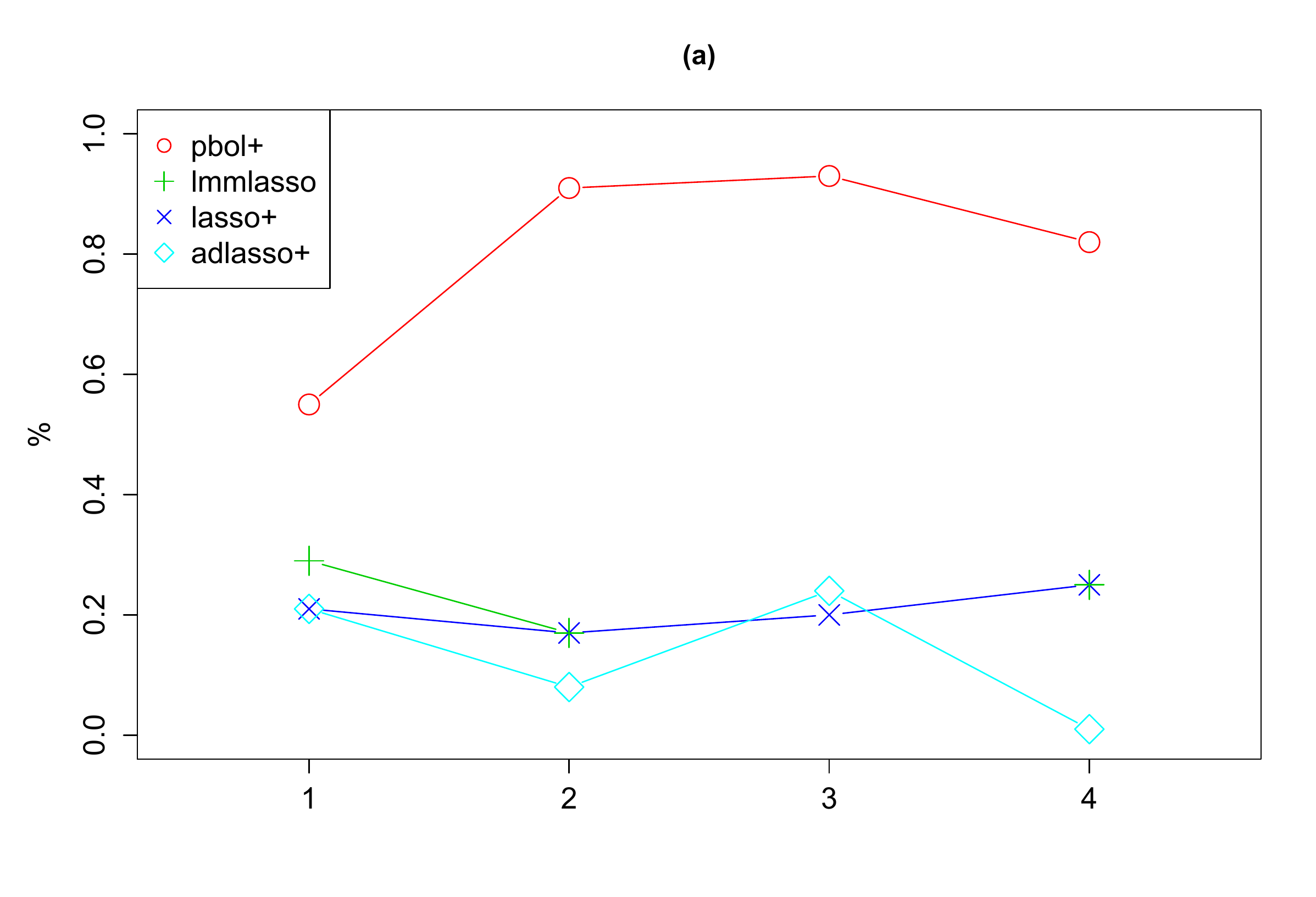}
\label{Truth}}
{\includegraphics[width=7.5cm]{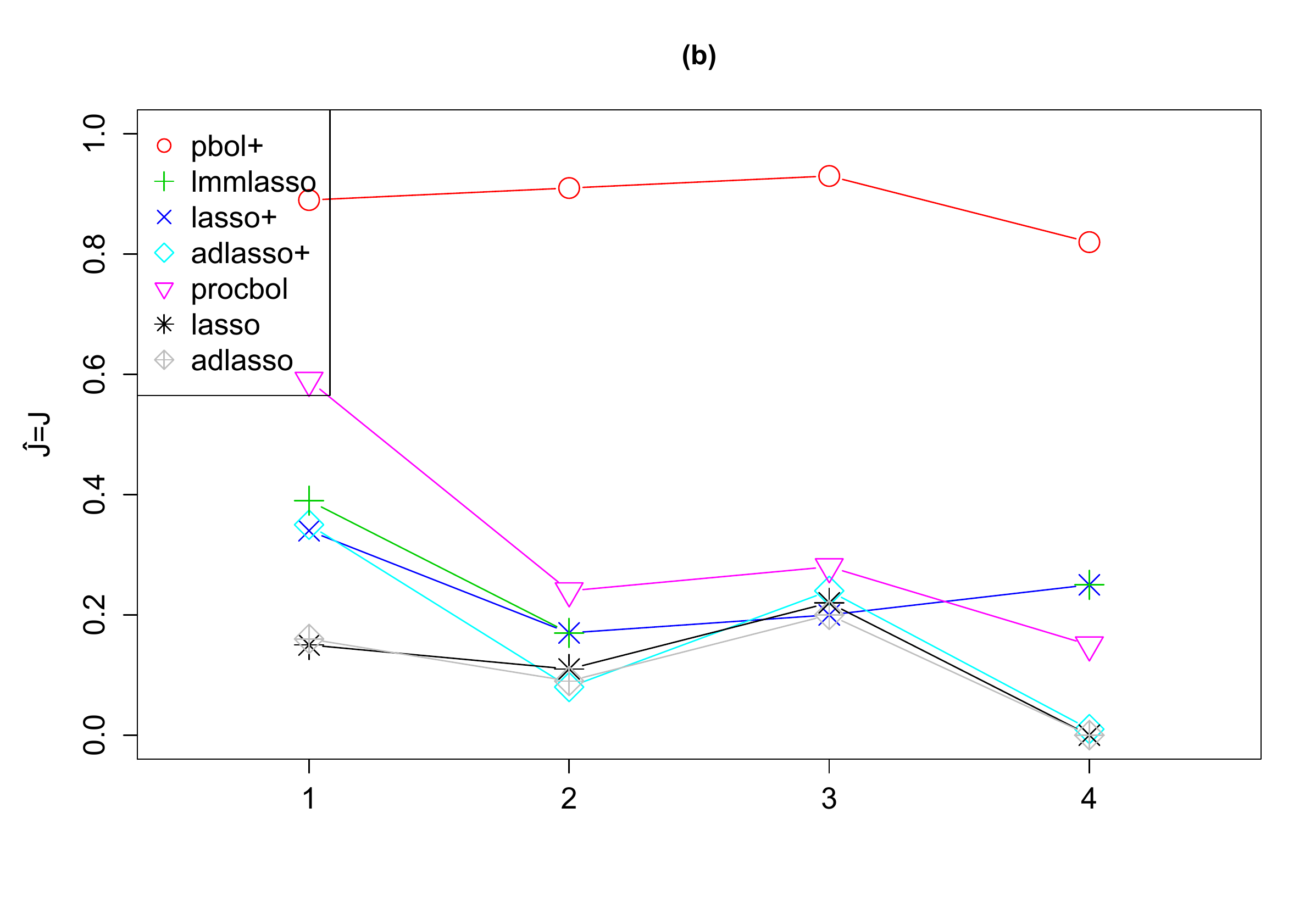}
\label{J}}
{\includegraphics[width=7.5cm]{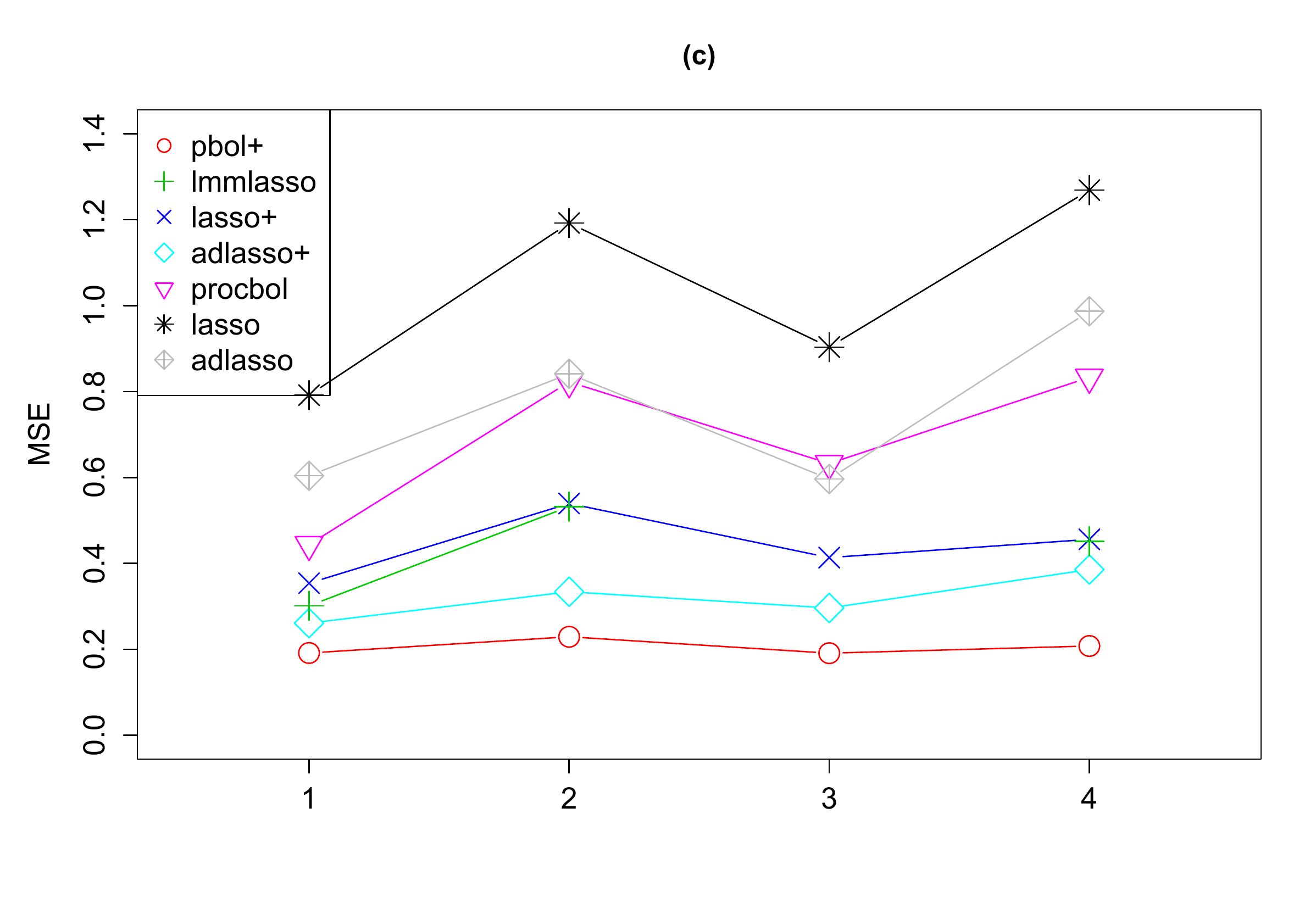}
\label{MSE}}
\caption{Summary of the results of the simulation study for models $M_1-M_4$ ($X$ axis). Results of `Truth'  (a), `$\hat J=J$' (b) and Mean Squared Error (c) for each model. }
\label{summary}
\end{figure}

The detailed results of the simulation study are available in Appendix A. A summary of the main results is shown in Figure \ref{summary} ($\alpha=0.1$ for the \textit{procbol} method and the \textit{pbol+} method). No results are given for the \textit{lmmLasso} of \citet{Schelldorfer:2011} in Model $M_3$ since two different grouping factors are considered and the R-package lmmLasso does not include that setting.\\

In all models, there is an improvement of the results when we switch from a simple linear model to a linear mixed model; indeed there is a significant difference between \textit{Lasso} and  \textit{Lasso+} or \textit{procbol} and \textit{pbol+}, especially with model $M_4$.

\begin{figure}
\centering
{\includegraphics[width=8cm]{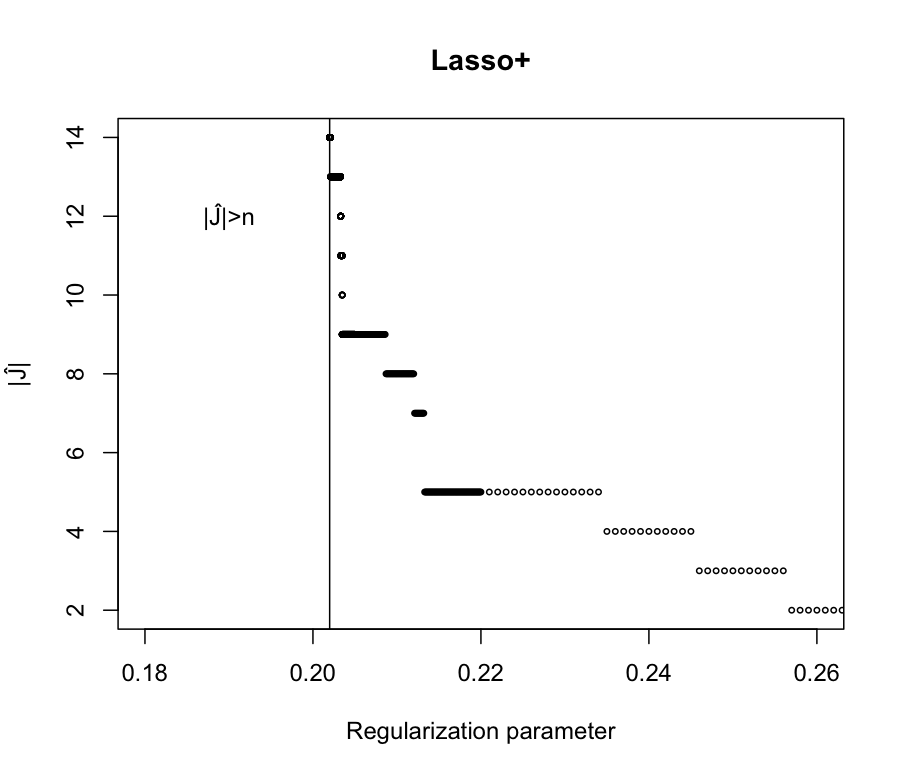}\label{gap_lasso}}
{\includegraphics[width=8cm]{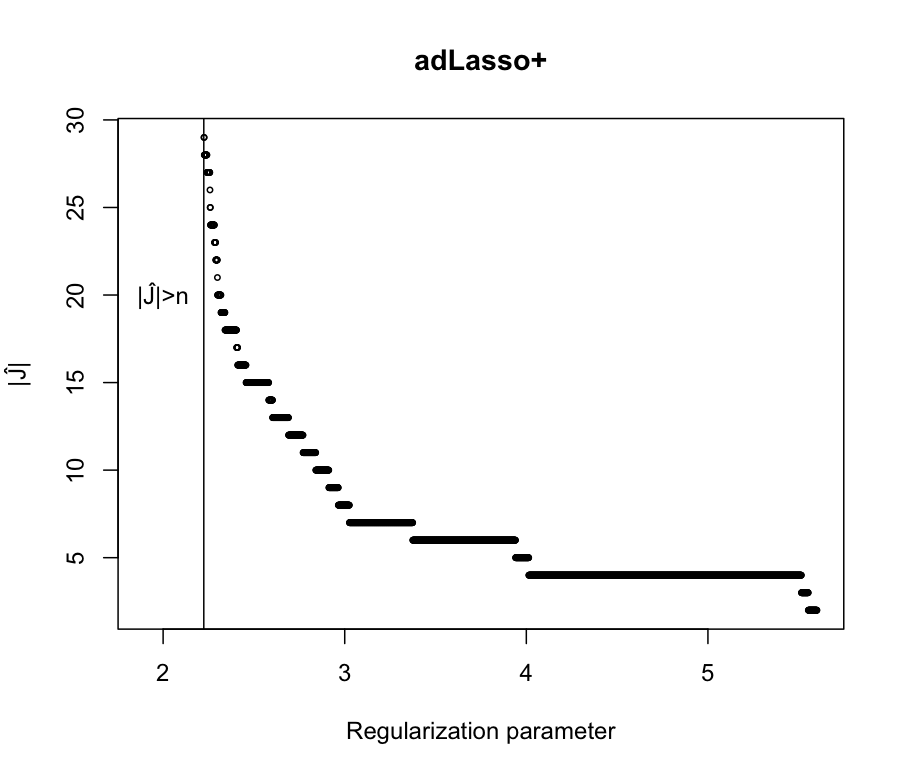}\label{gap_adlasso}}
{\caption{Number of selected fixed effects coefficients depending on the value of the regularization parameter for one run of model M4, for the method (a) \textit{Lasso+} and (b) \textit{adLasso+}. The grid of the penalty is as thin as $10^{-7}$ next to the area $|\hat J|>n$ in (a) and $10^{-3}$ in (b). }
\label{gap}
}
\end{figure}

On all models, \textit{lmmLasso} and \textit{Lasso+} give very similar results; this is not surprising since both are a $\ell^1$-penalization of the log likelihood, except for model $M_1$ where \textit{lmmLasso} seems to give better results. This difference comes from the coding of the R-package that contains the \textit{lmmLasso} method. Indeed, a variable that generates both a fixed and a random effect does not undergo feature selection in the \textit{lmmLasso} method when the random effect tends towards zero, whereas the \textit{Lasso+} method would allow it. \\
 We observed on our simulation study that both \textit{lmmLasso} and \textit{Lasso+} are very sensitive to the choice of the regularization parameter. On most simulations of model $M_4$ in which $p=600$, we observed an edge effect between a regularization parameter that selects few fixed effects (fewer than $15$) and a regularization parameter that selects too much fixed-effects  ($|\hat J|>n$) and thus stops the algorithm because we assumed that the number of relevant fixed-effects is lower than $\text{min}(n-1,p)$, see Figure \ref{gap_lasso}. Nevertheless, the weights included in the \textit{adLasso+} seems to smooth this phenomenon, see Figure \ref{gap_adlasso} for the same simulation as Figure \ref{gap_lasso}. Remark that for the run of model $M_4$ which is on Figure \ref{gap}, \textit{Lasso+} could select the true model for a regularization parameter around $0.22$ whereas \textit{adLasso+} could not as a noisy variable enters the set of selected variables before all the relevant fixed-effects do.
 
Concerning the \textit{adLasso+} method, it appears to improve the \textit{Lasso+} method, except for model $M_4$ where the true model is only selected once over the $100$ replications. On this particular model $M_4$, \textit{adLasso+} selects more fixed effects but less relevant ones than \textit{Lasso+}. This could mean that the initial weights are not adapted to this case. Despite the result of `Truth', the \textit{mse} is lower for \textit{adLasso+} than for \textit{Lasso+}.

Algorithm \ref{alg3} combined with the procbol method (\textit{pbol+}) gives the best results over all tested methods for all models. Indeed the percentage of true model recovered is the largest over all methods, the estimation of the fixed effects is really close to the reality and the $mse$ is the lowest among the tested methods. Nevertheless, due to the bias of the Lasso, the results in term of \textit{mse} for  \textit{Lasso+} and \textit{lmmLasso}  could easily be improved with a linear mixed model estimation as said in Section \ref{alg} (see Appendix). Yet, the results of \textit{pbol+} are mitigated for model $M_1$. Indeed, the percentage of true model recovered is lower than in the other models because of the selection of the random effects that lacks efficiency (the results concerning the selection of the fixed-effects are equivalent as in the other models, as shown in Figure \ref{summary}). Nonetheless, the results are still better than for the others methods. Moreover, a relevant random effect was never falsely deleted in all models and for all methods. It is interesting to note that the \textit{pbol+} method always converged on our simulations.

 A R-package ``MMS" is available 
 on CRAN (http://cran.r-project.org). This package contains tools to perform fixed effects selection in linear mixed models; it contains the previous methods denoted as  \textit{Lasso+}, \textit{adLasso+}, \textit{pbol+}, among others. \\
All the results presented in this section were obtained with a specific initialization of the algorithms. The next paragraph is dedicated to the analysis of the influence of that specific initialization.

\subsection{Influence of the initialization of our algorithms}
\label{influence}

Both Algorithm \ref{alg1} and Algorithm \ref{alg3} start with an initialization of the parameter $\Phi=(\sigma_1^2,\dots,\sigma_q^2,\sigma_e^{2},\beta)$. We choose to initialize each algorithm with the following setting: for all $1\le k\le q, \sigma_{k}^{2[0]}=\frac{0.4}{q} \ \sigma_e^{2[-1]}, \sigma_e^{2[0]}=0.6\ \sigma_e^{2[-1]}$, and $(\sigma_e^{2[-1]},\beta^{[0]})$ is estimated from a linear estimation (without the random effects) of the method used at Step 2.

In the current Section, we choose different initializations of Algorithm \ref{alg1} and Algorithm \ref{alg3}, both on Model $M_4$ (see Section \ref{results}). 
The initial values of the variances were set from $0.1$ to $10$ and of the fixed effects coefficients from $-100$ to $100$. Each algorithm always converged towards the same point, whatever the initialization of $\Phi$, not shown. However, the farther $\Phi^{[0]}$ is set from the true estimation of $\Phi$, the higher is the number of iterations of the algorithms.

\section{Application on a real data-set}
\label{real_data}
In this section we analyze a real data set which comes from \cite{Rohart:2012}. The aim of this analysis is to pinpoint metabolomic data that describes a phenotype taking into account all the available information such as the breed, the batch effect and the relationship between individuals. Here we will study the Daily Feed Intake phenotype (DFI). 
We model the data as follows:
\begin{equation}
y=X_B\beta_B+X_M\beta_M+Z_Eu_E+Z_Fu_F+\epsilon,
\end{equation}
where $y$ is the DFI phenotype, $X_B,X_M,Z_E, Z_F$ are the design matrices of the breed effect, the metabolomic data, the batch effect and the family effect, respectively. We consider two random effects: the batch and the family, considering that each level of these factors is a random sample drawn from a much larger population of batches and families, contrary to the breed factor. Note that the coefficients $\beta_B$ do not undergo feature selection.\\
We compare several methods on this model: \textit{Lasso}, \textit{adLasso}, \textit{procbol}, \textit{Lasso+}, \textit{adLasso+} and \textit{pbol+} (see Section \ref{results}). The model which is considered for the first three methods is $y=X_B\beta_B+X_M\beta_M+\epsilon$. Both methods \textit{procbol} and \textit{pbol+} were set with a user-level of $\alpha=0.1$. The results are presented in Table \ref{RD}.\\
\begin{table}
\centering
\begin{tabular}{l|c| c | c | c}
&$|\hat{J}|$& $\hat\sigma_e^2$ & $\hat\sigma_E^2$ &$\hat\sigma_F^2$\\
\hline
Lasso& 14&$3.8\times 10^{-2}$  &- &- \\
 adLasso&21 & $3.4\times 10^{-2}$&- &- \\
procbol & 11&$4.1 \times 10^{-2}$ &- &-\\
\hdashline
Lasso+ &$11$ &$3.2 \times 10^{-2}$ &$3.2 \times 10^{-3}$ &$ 6.4\times 10^{-3}$ \\
adLasso+&$10 $&$3.3 \times 10^{-2}$ & $2.5 \times  10^{-3}$ &$6.5 \times  10^{-3}$ \\
pbol+ &5 &$3.4 \times 10^{-2}$ &$5.9 \times 10^{-3}$ &$ 6.5\times 10^{-3}$ \\
\end{tabular}
\caption{Results for the real data set}
\label{RD}
\end{table}
We observe that considering random effects leads to a decrease of both the residual variance and the number of selected metabolomic variables. This behavior is in accordance with the simulation study. The question that arises from this analysis is to know whether the variables which are selected in the linear mixed models are more relevant than in the linear model. Biological analyses remain to be done to answer that question.\\

Table \ref{time} gives the computational time of one run when we only consider the batch effect -in order to be able to compute the \textit{lmmLasso}-, showing that the \textit{Lasso+} method is much faster than the \textit{lmmLasso} method for a large number of observations (due to the inversion of the matrix of variance $V$ at each step of the convergence process). The simulation was performed at a regularization parameter that selects the same model for the two methods, on a $2.80$GHz CPU with $8.00$Go of RAM. 
\\
\begin{table}
\centering
\begin{tabular}{l|c}
Methods&CPU Time \\
\hline
Lasso+ &0.80 \\
lmmLasso&24.28 \\
\end{tabular}
\caption{CPU Time on a single run that selects the same model}
\label{time}
\end{table}

\section{Conclusion}
In this paper, we proposed to add a $\ell^1$-penalization of the complete log-likelihood in order to perform selection of the fixed effects in a linear mixed model. The multicycle ECM algorithm used to minimize the objective function also performs random effects selection. This algorithm gives the same results as the lmmLasso of \cite{Schelldorfer:2011} when the random effects are assumed to be independent, but faster. Theoretical results are identical to those of \cite{Schelldorfer:2011} when the variances are known. The structure of our algorithm gives the possibility to combine it with any variable selection method built for linear models, but at the price of possibly loosing the convergence property. Nonetheless, the combined procbol method appears to give good results on simulated data and outperforms other approaches.\\
We applied all these methods to a real data set showing that the residual variance can be reduced, even with a small set of selected variables.

\bibliography{biblioMM}

\newpage
\section*{Appendix A - Results of the simulation study}

\label{app1}

\begin{table}[h!]
\centering
\small\caption{ \small Results of model $M_1$. The percentage of true model recovered was recorded -`Truth'- as well as  $\hat J=J$. $|J|$ is the number of fixed effects selected and $TP$ the number of relevant fixed effects selected. The signal to noise ratio is equal to $SNR=0.78 (0.13)$. Standard errors are given in parentheses, for $100$ runs.}
\small
\begin{tabular}{l|c| c | c | c |c| c | c | c }
&Truth &$\hat J=J$&$|\hat{J}|$& TP & $\hat\sigma_e^2$ & $\hat\sigma_1^2$ &$\hat\sigma_2^2$ &$\hat\sigma_3^3$\\
\hline
Ideal &1 &5 &5 &5 &1 &1 &1 &0 \\
\hdashline
Lasso&- &0.15 &4.95 &4.13 &3.27 &- &- &- \\
& & &(1.90) &(1.12) &(0.62)&-&-&- \\
adLasso&- &0.16 &5.25 &4.26 &2.91 &- &- &- \\
& & &(1.84) &(0.89) &(0.59)&-&-&- \\
procbol &- &0.59 &4.70 &4.58 &2.83 &- &- &- \\
$\alpha=0.1$& & &(0.78) &(0.61) &(0.57)&-&-&- \\
procbol &- &0.45 &4.47 &4.40 &2.89 &- &- &- \\
$\alpha=0.05$& & &(0.67) &(0.62) &(0.58)&-&-&- \\
\hdashline
Lasso+ &0.21 &0.34 &6.42 &5.00 &1.04 &0.88 &0.98 &0.02 \\
& & &(1.64) &(0.00) &(0.21)&(0.37)&(0.44)&(0.06) \\
adLasso+&0.21 &0.35 &6.34 &4.99 &0.94 &0.86 &0.95 &0.02 \\
& & &(1.41) &(0.10) &(0.18)&(0.36)&(0.41)&(0.06) \\
lmmLasso&0.29 &0.39 &6.15 &5.00 &1.01 &0.89 &0.96 &0.02 \\
& & &(1.29) &(0.00) &(0.19)&(0.38)&(0.42)&(0.06) \\
pbol+ &0.55 &0.89 &5.18 &5.00 &0.92 &0.87 &0.97 &0.03 \\
$\alpha=0.1$& & &(0.50) &(0.00) &(0.18)&(0.37)&(0.41)&(0.06) \\
pbol+ &0.59 &0.93 &5.08 &5.00 &0.93 &0.88 &0.97 &0.03 \\
$\alpha=0.05$& & &(0.30) &(0.00) &(0.17)&(0.37)&(0.41)&(0.06) \\
\end{tabular}

\vspace{0.2cm}

\begin{tabular}{l|c| c | c | c |c| c }
&$\hat\beta_1$&$\hat\beta_2$&$\hat\beta_3$&$\hat\beta_4$&$\hat\beta_5$& MSE\\
\hline
Ideal&0.67 &0.67 &0.67 &0.67 &0.67 &0.00\\
\hdashline
Lasso &0.67 &0.29 &0.31 &0.41 &0.17 &0.79\\
&(0.27) &(0.26) &(0.20) &(0.19) &(0.16)&(0.42) \\
adLasso&0.69 &0.42 &0.46 &0.58 &0.27 &0.60\\
&(0.27) &(0.33) &(0.25) &(0.23) &(0.22)&(0.37) \\
procbol &0.69 &0.63 &0.68 &0.65 &0.49 &0.44\\
$\alpha=0.1$&(0.27) &(0.32) &(0.17) &(0.30) &(0.33)&(0.31) \\
procbol &0.69 &0.63 &0.68 &0.62 &0.43 &0.51\\
$\alpha=0.05$&(0.27) &(0.32) &(0.17) &(0.33) &(0.36)&(0.30) \\
\hdashline
Lasso+ &0.69 &0.65 & 0.49& 0.41&0.43 &0.35\\
&(0.25) &(0.28) &(0.17) &(0.11) &(0.11)&(0.17) \\
adLasso+& 0.69&0.64 &0.59 &0.57 &0.48 &0.26\\
&(0.25) &(0.27) &(0.15) &(0.12) &(0.14)&(0.15) \\
lmmLasso&0.69 &0.65 &0.66 &0.41 &0.43 &0.30\\
&(0.25) &(0.28) &(0.11) &(0.11) &(0.10)&(0.15) \\

pbol+ &0.69 &0.67 &0.67 &0.66 &0.65 &0.19\\
$\alpha=0.1$&(0.25) &(0.28) &(0.12) &(0.10) &(0.10)&(0.14) \\
pbol+ &0.69 &0.67 &0.67 &0.66 &0.65 &0.18\\
$\alpha=0.05$&(0.25) &(0.28) &(0.11) &(0.10) &(0.10)&(0.13) \\
\end{tabular}
\label{M1}
\end{table}

\newpage
\begin{table}[h!]
\centering
\caption{\small Results of model $M_2$. The percentage of true model recovered was recorded -`Truth'- as well as  $\hat J=J$. $|J|$ is the number of fixed effects selected and $TP$ the number of relevant fixed effects selected. The signal to noise ratio is equal to $SNR=1.02 (0.21)$. Standard errors are given in parentheses, for $100$ runs.}
\small
\begin{tabular}{l|c| c | c | c |c| c | c }
Results &Truth &$\hat J=J$&$|\hat{J}|$& TP & $\hat\sigma_e^2$ & $\hat\sigma_1^2$ &$\hat\sigma_2^2$ \\
\hline
Ideal &1 &5 &5 &5 &1 &1 &1  \\
\hdashline
Lasso & - &0.11  &5.02  &3.86  &3.62  & -&-\\
&&&(2.69) &(1.35) &(0.96)  &-&- \\
adLasso& -&0.09 &6.06  &4.24  &3.05  &-&- \\
&&  &(2.66) &(1.16) &(0.87)&- &-\\
procbol  &- &0.24  &3.95  &3.76  &3.62 &-  &-  \\
$\alpha=0.1$&  &&(1.22) &(1.06) &(0.95)&-&- \\
procbol& -&0.21 &3.60  &3.47  &3.53  &- &-\\
$\alpha=0.05$&&  &(1.25) &(1.14) &(0.87)&- &-\\
\hdashline
Lasso+ &0.17 &0.17  &7.60  &4.92  &1.25  &0.91&0.93  \\
&  &&(2.64) &(0.37) &(0.28)&(0.40)&(0.48) \\
adLasso+ &0.08 &0.08  &8.26  &5.00  &0.99  &0.90&0.85  \\
&  &&(3.15) &(0.00) &(0.21)&(0.38)&(0.41) \\
lmmLasso &0.17 & 0.17 &7.65  &4.93  &1.24  &0.91&(0.93)  \\
&  &&(2.49) &(0.36) &(0.26)&(0.40) &(0.48)\\
pbol+  &0.91 &0.91  & 4.86 & 4.85 &1.01  &0.95&0.88  \\
$\alpha=0.1$&  &&(0.59) &(0.58) &(0.28)&(0.38)&(0.41) \\
pbol+  &0.80 &0.80  &4.57  &4.57  &1.11  &0.93 &0.88 \\
$\alpha=0.05$&  &&(0.93) &(0.93) &(0.39)&(0.38) &(0.39)\\
\end{tabular}

\vspace{0.2cm}

\begin{tabular}{l|c| c | c | c |c| c }
&$\hat\beta_{i_1}$&$\hat\beta_{i_2}$&$\hat\beta_{i_3}$&$\hat\beta_{i_4}$&$\hat\beta_{i_5}$& MSE\\
\hline
Ideal& 0.75 &0.75  &0.75  &0.75  &0.75  &0.00 \\
\hdashline
Lasso & 0.79 & 0.47 &0.21  &0.19  &0.17  &1.19 \\
&(0.27) &(0.31) &(0.19) &(0.17) &(0.16)&(0.57) \\
adLasso & 0.79 & 0.64 &0.36  &0.35  &0.29  &0.84 \\
&(0.27) &(0.38) &(0.24) &(0.24) &(0.22)&(0.55) \\
procbol  &0.79  &0.72  &0.50  &0.57  &0.52  &0.82 \\
$\alpha=0.1$&(0.27) &(0.49) &(0.40) &(0.38) &(0.38)&(0.55) \\
procbol  &0.79  &0.75  &0.44  &0.50  &0.45  &0.93 \\
$\alpha=0.05$&(0.27) &(0.50) &(0.42) &(0.41) &(0.40)&(0.56) \\
\hdashline
Lasso+ &0.82  &0.91  &0.35  &0.35  &0.33  &0.54 \\
&(0.26) &(0.26) &(0.13) &(0.11) &(0.13)&(0.24) \\
adLasso+  &0.81  &0.82  & 0.51 & 0.52 &0.49  &0.33 \\
&(0.25) &(0.25) &(0.14) &(0.13) &(0.14)&(0.17) \\
lmmLasso &0.82  &0.91  &0.35  &0.35  &0.33  &0.53 \\
&(0.26) &(0.26) &(0.13) &(0.11) &(0.13)&(0.23) \\
pbol+  &0.79  &0.76  &0.70  &0.73  &0.72  &0.23 \\
$\alpha=0.1$&(0.25) &(0.26) &(0.22) &(0.17) &(0.18)&(0.28) \\
pbol+  &0.80  &0.79  &0.64  &0.66  &0.66  &0.35 \\
$\alpha=0.05$&(0.25) &(0.28) &(0.29) &(0.28) &(0.28)&(0.43) \\
\end{tabular}
\label{M2}
\end{table}

\newpage
\begin{table}[h!]
\centering
\caption{\small Results of model $M_3$. The percentage of true model recovered was recorded -`Truth'- as well as  $\hat J=J$. $|J|$ is the number of fixed effects selected and $TP$ the number of relevant fixed effects selected. The signal to noise ratio is equal to $SNR= 0.83(0.16)$. Standard errors are given in parentheses, for $100$ runs.}
\small
\begin{tabular}{l|c| c | c | c |c| c | c }
Results &Truth &$\hat J=J$&$|\hat{J}|$& TP & $\hat\sigma_e^2$ & $\hat\sigma_1^2$ &$\hat\sigma_2^2$ \\
\hline
Ideal &1 &5 &5 &5 &1 &1&1  \\
\hdashline
Lasso &- &0.22 &4.96 &4.13 &3.32 & -&-  \\
&  & &(2.18) &(1.10) &(0.80)&-&- \\
adLasso&- &0.20 &6.10 &4.58  &2.85 & -&- \\
&&  &(2.19) &(0.70) &(0.72)&- &-\\
procbol &-&0.28 &4.37 &4.12 &2.90 &- &-  \\
$\alpha=0.1$& & &(1.08) &(0.77) &(0.79)&-&- \\
procbol &-&0.26 &4.17 &3.97 &2.97 &- &-  \\
$\alpha=0.05$&  &&(1.12) &(0.83) &(0.82)&-&- \\
\hdashline
Lasso+ &0.20&0.20 &7.07 &4.99 &1.11 &0.91 &0.92  \\
&  &&(2.01) &(0.10) &(0.22)&(0.36)&(0.46) \\
adLasso+&0.24 &0.24 &6.70 &4.97  &0.97 &0.88 &0.88 \\
& & &(1.51) &(0.17) &(0.19)&(0.34)&(0.45) \\
lmmLasso&-&- &- &- & -& -&- \\
&  &&- &- &-&-&- \\
pbol+ & 0.93& 0.93&5.09 &5.00 &0.95 &0.91 &0.89  \\
$\alpha=0.1$&  &&(0.38) &(0.00) &(0.17)&(0.33)&(0.44) \\
pbol+ &0.95&0.95 &5.08 &5.00 &0.95 &0.91 &0.89 \\
$\alpha=0.05$&  &&(0.44) &(0.00) &(0.17)&(0.33)&(0.44) \\
\end{tabular}

\vspace{0.2cm}

\begin{tabular}{l|c| c | c | c |c| c }
&$\hat\beta_1$&$\hat\beta_2$&$\hat\beta_3$&$\hat\beta_4$&$\hat\beta_5$& MSE\\
\hline
Ideal&0.67 &0.67 &0.67 &0.67 &0.67 &0.00\\
\hdashline
Lasso &0.69 &0.69 &0.18 &0.20 &0.27 &0.90\\
&(0.25) &(0.32) &(0.17) &(0.17) &(0.17)&(0.40) \\
adLasso&0.69 &0.68 &0.32 &0.36 &0.46 &0.60\\
&(0.25) &(0.32) &(0.21) &(0.21) &(0.22)&(0.32) \\
procbol &0.73 &0.65 &0.48 &0.51 &0.57 &0.63\\
$\alpha=0.1$&(0.34) &(0.13) &(0.36) &(0.36) &(0.35)&(0.42) \\
procbol &0.73 &0.65 &0.44 &0.49 &0.56 &0.68\\
$\alpha=0.05$&(0.34) &(0.13) &(0.38) &(0.38) &(0.36)&(0.43) \\
\hdashline
Lasso+ &0.71 &0.71 &0.40 &0.38 &0.43 &0.41\\
&(0.24) &(0.29) &(0.12) &(0.11) &(0.11)&(0.19) \\
adLasso+&0.71 &0.69 &0.50 &0.48 &0.56 &0.30\\
&(0.24) &(0.29) &(0.16) &(0.14) &(0.13)&(0.18) \\
lmmLasso&- &- &- &- &- &-\\
&- &- &- &- &-&- \\
pbol+ &0.71 &0.69 &0.67 &0.65 &0.68 &0.19\\
$\alpha=0.1$&(0.24) &(0.29) &(0.12) &(0.10) &(0.10)&(0.16) \\
pbol+ &0.71 &0.69 &0.67 &0.65 &0.68 &0.19\\
$\alpha=0.05$&(0.24) &(0.29) &(0.12) &(0.10) &(0.10)&(0.16) \\
\end{tabular}
\label{M3}
\end{table}

\newpage
\begin{table}[h!]
\centering
\caption{\small Results of model $M_4$. The percentage of true model recovered was recorded -`Truth'- as well as  $\hat J=J$. $|J|$ is the number of fixed effects selected and $TP$ the number of relevant fixed effects selected. The signal to noise ratio is equal to $SNR=0.63 (0.11)$. Standard errors are given in parentheses, for $100$ runs.}
\small
\begin{tabular}{l|c| c | c | c |c| c | c }
Results &Truth &$\hat J=J$&$|\hat{J}|$& TP & $\hat\sigma_e^2$ & $\hat\sigma_1^2$ &$\hat\sigma_2^2$ \\
\hline
Ideal &1 &5 &5 &5 &1 &1 &1 \\
\hdashline
Lasso &- &0.00 &2.81 &2.06 &4.08 &- &-  \\
&&  &(2.80) &(1.30) &(0.84)&-&- \\
adLasso&- &0.00 &5.64 &3.03 &3.38 &-&- \\
&  & &(4.10) &(1.22) &(0.88)&-&- \\
procbol &- &0.15 &3.85 &3.61 &3.23 &- &-  \\
$\alpha=0.1$&  &&(1.00) &(0.95) &(0.73)&-&- \\
procbol &-&0.15 &3.48 &3.34 &3.39 &- &- \\
$\alpha=0.05$&&  &(1.00) &(0.99) &(0.80)&-&- \\
\hdashline
Lasso+ &0.25&0.25 &7.13 &4.99 &1.21 &0.93 &1.03\\
&  &&(1.84) &(0.10) &(0.27)&(0.41) &(0.40)\\
adLasso+&0.01 &0.01 &9.56 &4.87 &0.94 &0.89 &0.98  \\
&  &&(4.01) &(0.37) &(0.26)&(0.37)&(0.37) \\
lmmLasso&0.25&0.25 &7.22 &4.99 &1.19 &0.93 &1.03 \\
& & &(1.95) &(0.10) &(0.25)&(0.40)&(0.40) \\
pbol+ &0.82&0.82 &5.21 &4.99 &0.92 &0.97 &1.00 \\
$\alpha=0.1$&  &&(0.56) &(0.10) &(0.17)&(0.39)&(0.34) \\
pbol+ &0.88&0.88 &5.10 &4.98 &0.93 &0.97 &1.00 \\
$\alpha=0.05$&  &&(0.41) &(0.14) &(0.16)&(0.39)&(0.34) \\
\end{tabular}

\vspace{0.2cm}

\begin{tabular}{l|c| c | c | c |c| c }
&$\hat\beta_1$&$\hat\beta_2$&$\hat\beta_3$&$\hat\beta_4$&$\hat\beta_5$& MSE\\
\hline
Ideal&0.67 &0.67 &0.67 &0.67 &0.67 &0.00\\
\hdashline
Lasso &0.60 &0.06 &0.06 &0.06 &0.11 &1.27\\
&(0.25) &(0.15) &(0.11) &(0.11) &(0.15)&(0.32) \\
adLasso&0.60 &0.15 &0.18 &0.17 &0.26 &0.99\\
&(0.25) &(0.26) &(0.19) &(0.19) &(0.23)&(0.33) \\
procbol &0.60 &0.55 &0.38 &0.44 &0.43 &0.83\\
$\alpha=0.1$&(0.25) &(0.31) &(0.38) &(0.39) &(0.40)&(0.43) \\
procbol &0.60 &0.53 &0.32 &0.38 &0.35 &0.91\\
$\alpha=0.05$&(0.25) &(0.32) &(0.38) &(0.40) &(0.41)&(0.39) \\
\hdashline
Lasso+ &0.62 & 0.55& 0.31 &0.35 &0.37 &0.46 \\
&(0.25) &(0.27) &(0.11) &(0.12) &(0.12)&(0.20) \\
adLasso+&0.61 &0.56 &0.41 &0.43 &0.48 &0.39\\
&(0.25) &(0.26) &(0.16) &(0.18) &(0.16)&(0.19) \\
lmmLasso&0.62 &0.55 & 0.31&0.35 &0.38 &0.45\\
&(0.25) &(0.27) &(0.11) &(0.12) &(0.12)&(0.19) \\
pbol+ &0.60 &0.64 &0.67 &0.67 &0.67 &0.21\\
$\alpha=0.1$&(0.25) &(0.28) &(0.10) &(0.11) &(0.13)&(0.15) \\
pbol+ &0.60 &0.64 &0.67 &0.67 &0.67 &0.20\\
$\alpha=0.05$&(0.25) &(0.27) &(0.10) &(0.13) &(0.13)&(0.15) \\
\end{tabular}
\label{M4}
\end{table}

\newpage
\begin{table}[h!]
\caption{\small Results of model $M_4$ when a ML linear regression is added after the convergence of the algorithm. The percentage of true model recovered was recorded -`Truth'- as well as  $\hat J=J$. $|J|$ is the number of fixed effects selected and $TP$ the number of relevant fixed effects selected. The signal to noise ratio is equal to $SNR=0.63 (0.11)$. Standard errors are given in parentheses, for $100$ runs.}
\centering
\small
\begin{tabular}{ c| c |*{2}{c}}
& Ideal &lmmLasso& Lasso+     \\
\hline
\hline
Truth 		& 1 		&0.25	&0.25		  \\
$\hat J=J$ &1&0.25&0.25\\
$|\hat{J}|$ &5		&7.22(1.95)	&7.13(1.84)  	 \\
$TP$ 		& 5 		&4.99(0.10)	&4.99(0.10)  	 \\
\hline
$\hat\sigma_e^2$ & 1		&1.19(0.25)	&1.21(0.27)  \\
$\hat\sigma_1^2$ & 1    		&0.96(0.39)	&0.96(0.40)    \\
$\hat\sigma_2^2$ & 1		&1.01(0.36)	&1.01(0.36)	\\
\hline
$\hat\beta_1$& 0.67		&0.61(0.25)	&0.61(0.25)  		  \\
$\hat\beta_2$& 0.67		&0.62(0.28)	&0.62(0.28)  	  \\
$\hat\beta_3$& 0.67		&0.61(0.12)	&0.61(0.12)  	\\
$\hat\beta_4$& 0.67		&0.63(0.12)	&0.63(0.12)  	\\
$\hat\beta_5$ & 0.67		&0.62(0.14)	&0.62(0.14)  	 \\
\hline
$mse$& 0 		&0.40(0.17)	&0.40(0.17) \\
\\\end{tabular}
\label{M4_REG}
\end{table}

\newpage
\section*{Appendix B - Remark on the tuning parameter}
\label{tun_lambda}
The tuning of the regularization parameter could be tricky for some methods, especially the \textit{Lasso} method and the \textit{adLasso} method. In this section, we look at the causes.

\begin{figure}[h!]
\centering
\subfigure[BIC or EBIC depending on the value of the regularization parameter of the Lasso method]{\includegraphics[width=6.5cm]{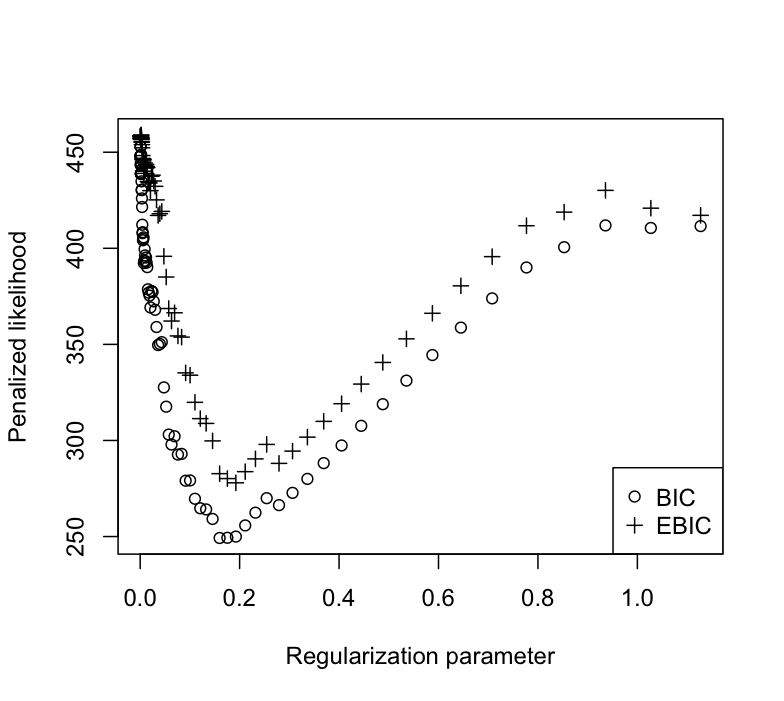}
\label{BIC-EBIC2}}\\
\subfigure[$-2\times$log-Likelihood depending on the regularization parameter of the Lasso method]{\includegraphics[width=6.5cm]{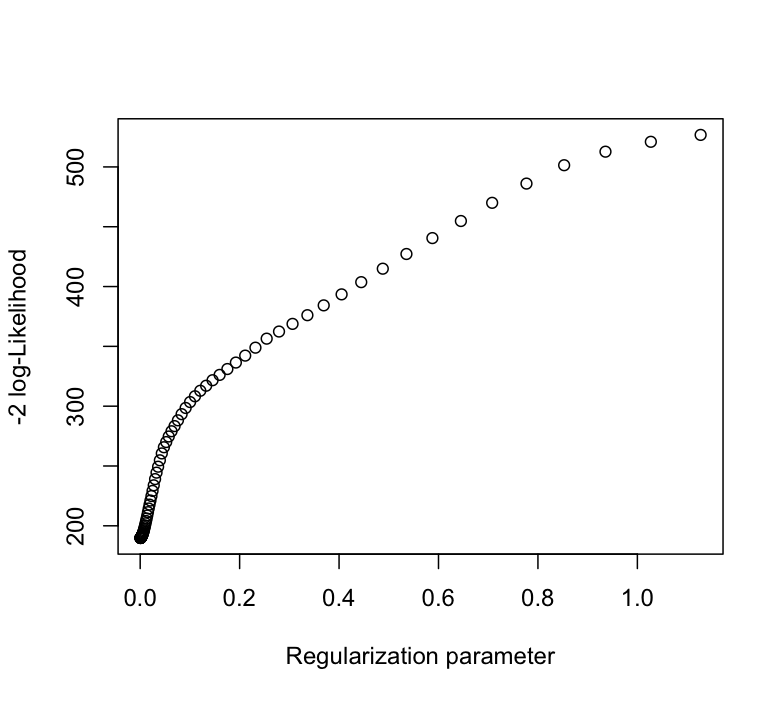}
\label{loglik2}}
\hspace{0.5cm}
\subfigure[Residual variance depending on the regularization parameter of the Lasso method]{\includegraphics[width=6.5cm]{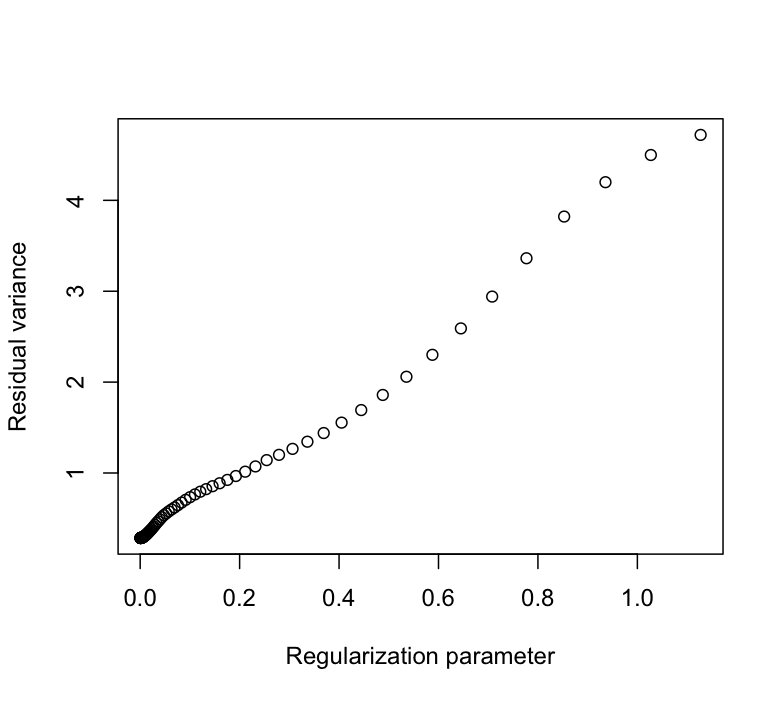}
\label{sigma_u2}}\\
\textbf{\caption{One simulation of linear model for the Lasso method with $n=120,p=80$ and $\beta_J=1$. }
\label{lowdim}
}
\end{figure}

We shall begin to consider the classical linear model before studying the linear mixed model.
Let us first look at the Lasso method when only applied in a classical linear model. We compare two penalizations of the likelihood: BIC and the Extended BIC (EBIC) \citep{Chen:2008}. The EBIC penalizes a space of dimension $k$ with a term that depends on the number of spaces that have the same dimension, which is $\frac{p!}{k!(p-k)!}$
; thus EBIC penalizes more the complex spaces than BIC. 
Figure \ref{lowdim} shows the behavior of the BIC and EBIC criteria, the log-likelihood and the residual variance for several values of the regularization parameter of the Lasso in a low dimensional case ($p=80$).
We observe that tuning the regularization parameter in this case raises no problem.

Let us now consider a simulation in a high dimensional context in which we have $n=120$ observations and $p=600$ explanatory variables. 
Results of the two methods for choosing the regularization parameter of Lasso are presented in Figure \ref{BIC}. 
\begin{figure}[h!]
\centering
\subfigure[BIC or EBIC depending on the value of the regularization parameter of the Lasso]{\includegraphics[width=6.5cm]{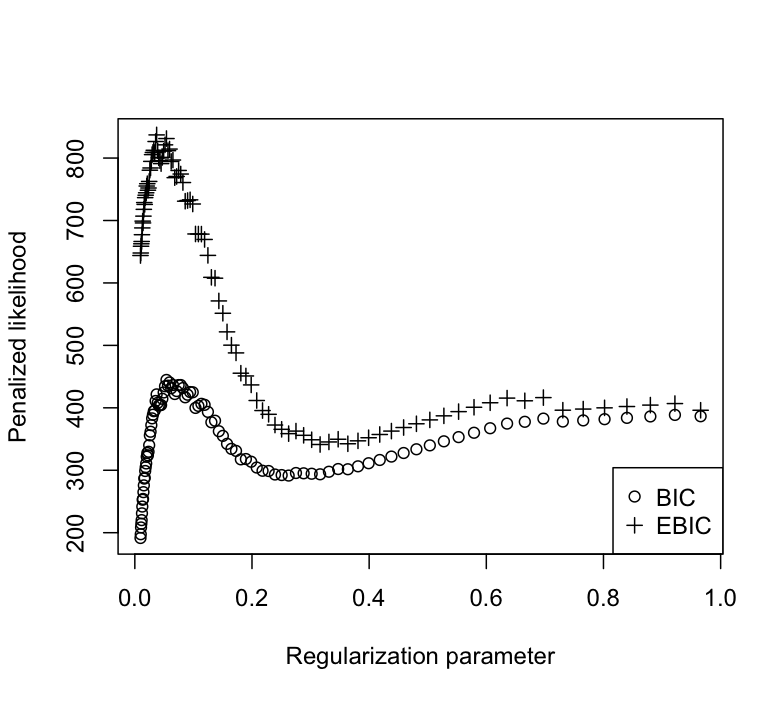}
\label{BIC-EBIC}}\\
\subfigure[$-2\times$log-Likelihood depending on the regularization parameter of the Lasso method]{\includegraphics[width=6.5cm]{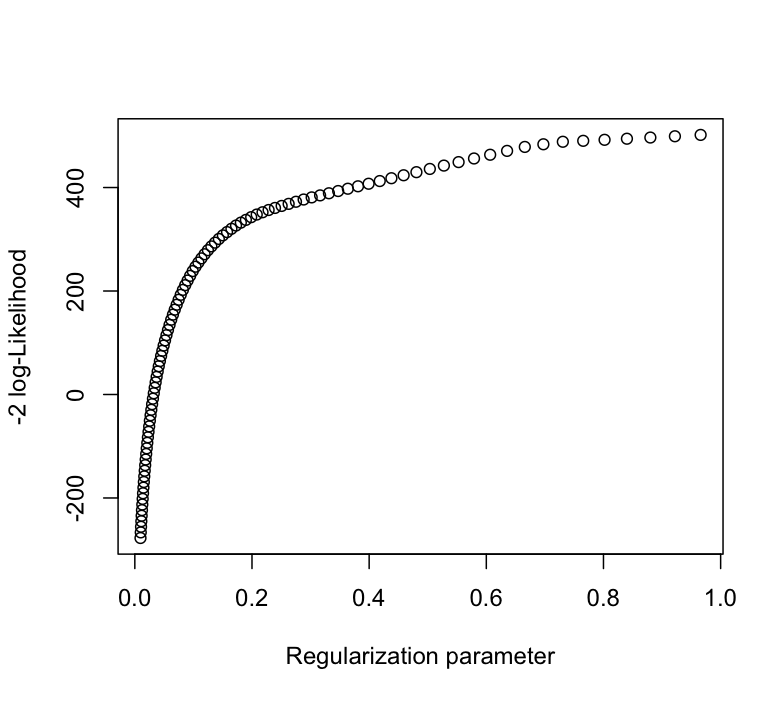}
\label{loglik1}}
\hspace{0.5cm}
\subfigure[Residual variance depending on the regularization parameter of the Lasso method]{\includegraphics[width=6.5cm]{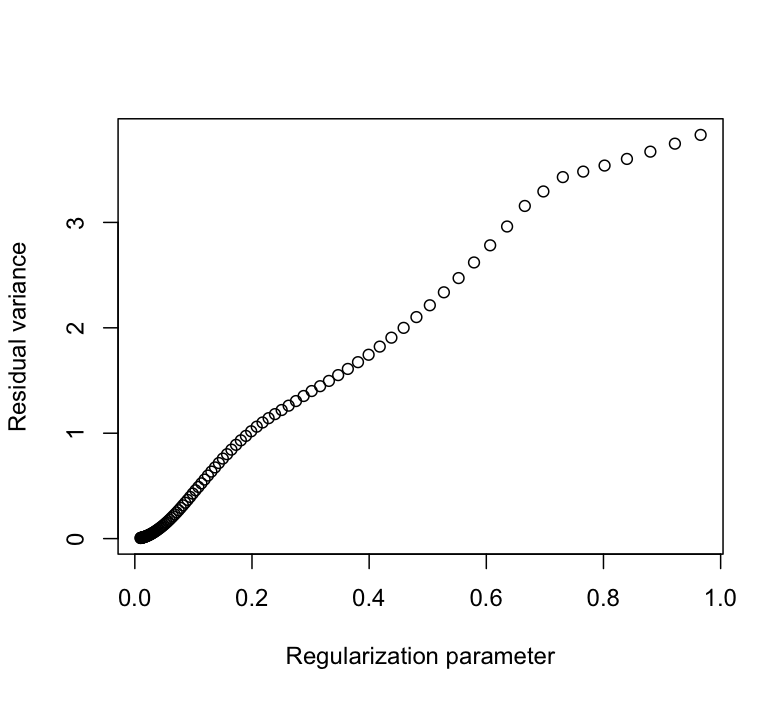}
\label{sigma_e}}\\
\textbf{\caption{One simulation of linear model for the Lasso method with $n=120,p=600$ and $\beta_J=1$. }
\label{BIC}
}
\end{figure}

Firstly, we confirm that EBIC is more conservative than BIC and penalizes more the complex spaces. 
On the far left of Figure \ref{BIC-EBIC}, we observe that both the BIC and the EBIC curves decrease when the regularization parameter is close to zero. This phenomenon is due to the degeneracy of the likelihood that can be seen in Figure \ref{loglik1} (stated in Section \ref{sec1} for mixed models, it can also happen in linear models). Figure \ref{sigma_e} shows that the degeneracy of the likelihood comes from the residual variance that drops to zero when the regularization parameter is close to zero, and thus when too much variables enter the model.

To conclude, we see that both BIC and EBIC penalties are not sufficiently strong to completely balance the degeneracy of the likelihood; however, EBIC penalty leads to select a more parsimonious model while BIC penalty selects a more complex model. Nonetheless, the EBIC penalty is usually too much conservative in practice, that is why the simulation study used the BIC penalty. 
When the degeneracy happens -as it is likely to occur as $p$ grows-, the regularization parameter should be optimized over an area that does not contain the explosion of the likelihood, that means that the area should not contain the far left part of  Figure \ref{BIC-EBIC} where the criterion decreases.\\

\begin{figure}[h!]
\centering
\subfigure[BIC or EBIC depending on the value of the regularization parameter of the Lasso+ method]{\includegraphics[width=6.5cm]{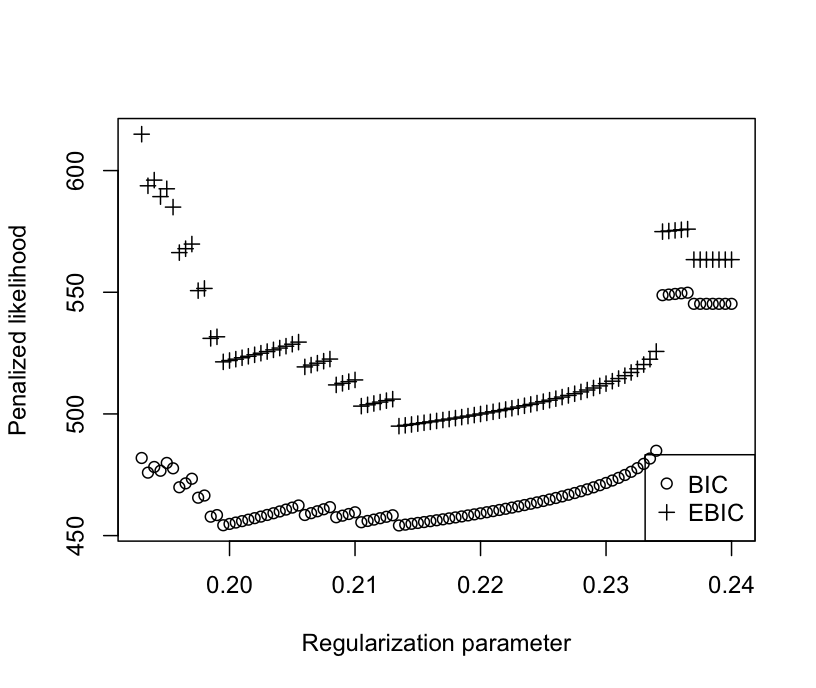}
\label{BIC-EBIC_MM}}
\hspace{0.5cm}
\subfigure[$-2\times$log-Likelihood depending on the regularization parameter of the Lasso+ method]{\includegraphics[width=6.5cm]{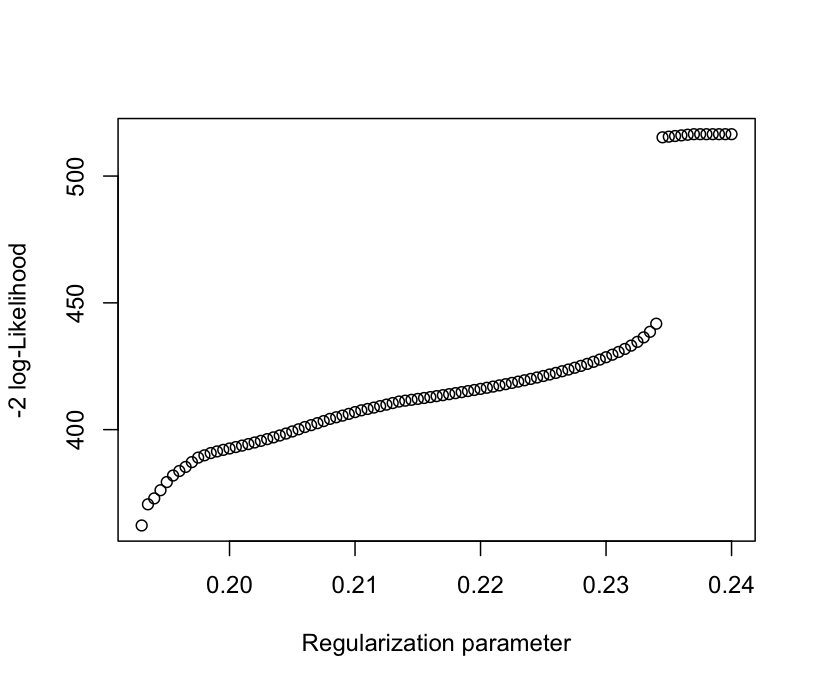}
\label{loglik_MM}}\\
\subfigure[Residual variance depending on the regularization parameter of the Lasso+ method]{\includegraphics[width=6.5cm]{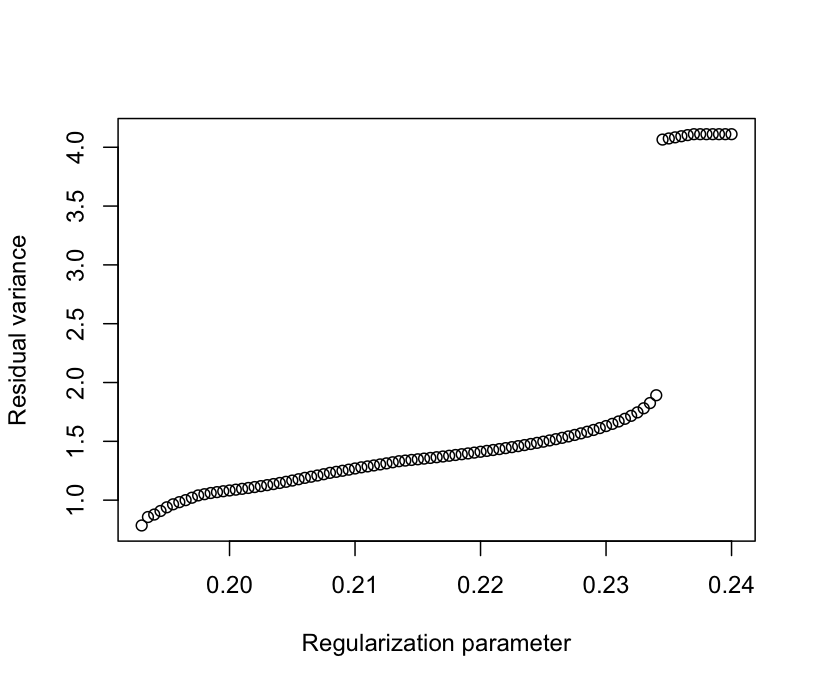}
\label{sigma_e_MM}}
\hspace{0.5cm}
\subfigure[Residual variance depending on the regularization parameter of the Lasso+ method]{\includegraphics[width=6.5cm]{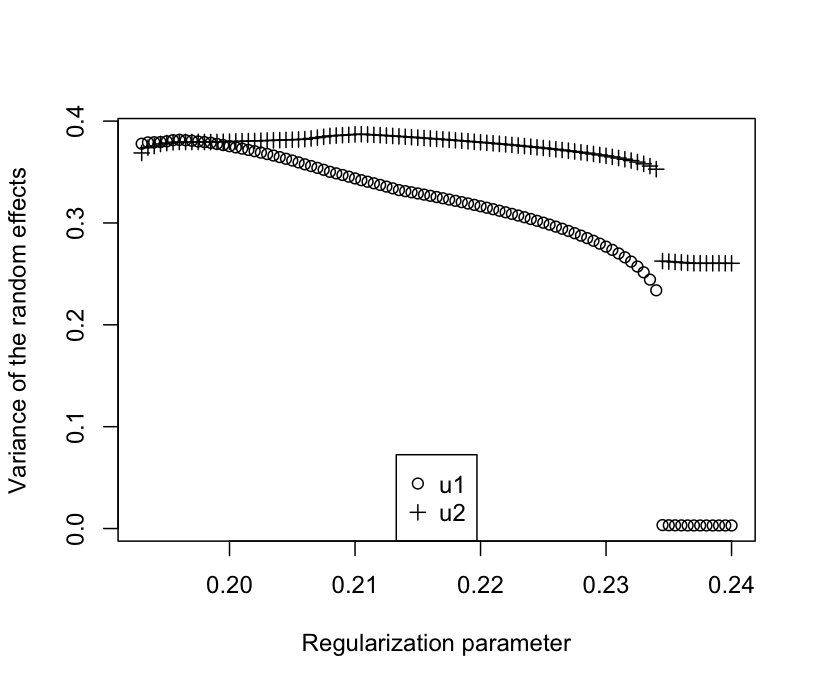}
\label{sigma_u_MM}}\\
\textbf{\caption{One simulation of linear mixed model with $n=120,p=600, \beta_J=1$ and two i.i.d. random effects. }
\label{BIC_MM}
}
\end{figure}

We now look at the \textit{Lasso+} method. As mentioned in the paper, the maximal number of fixed-effects that can be selected with the \textit{Lasso+} method is small compared to $n$ or $p$. Thus, the degeneracy of the likelihood  never occurred in our simulations (Figure \ref{BIC_MM}). However, if this phenomenon happens, the choice of the grid of the regularization parameter should follow the same advice as the one given above for the classical linear model.

\newpage
\section*{Appendix C - Proof of Proposition 2.2}
\label{proof}
$G$ and $R$ are supposed to be known. Thus the minimization of our objective function $g$ reduces to the minimization of the following function in $(\beta,u)$:\\
$h(u,\beta)=(y-X \beta-Z u)'R^{-1}(y-X \beta-Z u)+u'G^{-1}u+\lambda|\beta|_1$.\\
Let denote $(\hat u ,\hat\beta)=\underset{(u,\beta)}{\text{argmin } } h(u,\beta)$. Since the function $h$ is convex, we have:\\
$(\hat u ,\hat\beta)=\{\begin{array}{l l}
  u(\beta)=\underset{u}{\text{argmin } }h(u,\beta)\\
 \hat \beta=\underset{\beta}{\text{argmin } }h(u(\beta),\beta)\\
  \hat u =u(\hat \beta)\\
 \end{array}\right. $.\\
 Since $\dfrac{\partial h(u,\beta)}{\partial u} $ exists, we can explicit the minimum of $h$ in $u$:\\  $(\hat u ,\hat\beta)=
 \{\begin{array}{l l}
  u(\beta)=(Z'R^{-1}Z+G^{-1})^{-1}Z'R^{-1}(y-X \beta)\\
 \hat \beta=\underset{\beta}{\text{argmin } }h(u(\beta),\beta)\\
  \hat u =u(\hat \beta)\\
\end{array}\right. $\\
Thus, we obtain:\\

\begin{eqnarray*}
h(u(\beta), \beta)&=&(y-X \beta-Zu(\beta))'R^{-1}(y-X \beta-Zu(\beta))+u'G^{-1}u+\lambda| \beta|_1\\
&=&(y-X \beta)'R^{-1}(y-X \beta)-(y-X \beta)R^{-1}Zu(\beta)-(Zu(\beta))'R^{-1}(y-X \beta)\\
&&+(Z\hat u)'R^{-1}Zu(\beta)+u(\beta)'G^{-1}u(\beta)+\lambda| \beta|_1\\
&=&(y-X \beta)'\[R^{-1}-R^{-1}Z(Z'R^{-1}Z+G^{-1})^{-1}Z'R^{-1}\](y-X \beta)+\lambda| \beta|_1
\end{eqnarray*}
Denote $W=R^{-1}-R^{-1}Z(Z'R^{-1}Z+G^{-1})^{-1}Z'R^{-1}$. We can show that $W=(Z'GZ+R^{-1})^{-1}=V^{-1}$. 
This result comes from the equivalence between the resolution of Henderson's equations \citep{Henderson:1973} and the generalized least squares.

To conclude, we have that\\
$(\hat u ,\hat\beta)=\left((Z'R^{-1}Z+G^{-1})^{-1}Z'R^{-1}(y-X\hat\beta),\underset{\beta}{\text{argmin } }(y-X\beta)'V^{-1}(y-X\beta)+\lambda|\beta|_1\right)$.

\end{document}